# Developing Design Guidelines for Precision Oncology Reports

Selim Kalaycı*

Department of Genetics and
Genomics and Icahn Institute for
Genomics and Multi-scale Biology,
Icahn School of Medicine at Mount Sinai

Çağatay Demiralp†

MIT CSAIL & Fitnescity Labs

Zeynep H. Gümüş‡

Department of Genetics and
Genomics and Icahn Institute for
Genomics and Multi-scale Biology,
Icahn School of Medicine at Mount Sinai

## ABSTRACT

Precision oncology tests that profile tumors to identify clinically actionable targets have rapidly entered clinical practice. Effective visual presentation of the results of these tests is crucial in accurate clinical decision-making. In current practice, these results are typically delivered to oncologists as static prints, who then incorporate them into their clinical decision-making process. However, due to a lack of guidelines for standardization, different vendors use different report formats. There is very little known on the effectiveness of these report formats or the criteria necessary to improve them. In this study, we have aimed to identify both the tasks and the needs of oncologists from precision oncology report design and then to improve the designs based on these findings. To this end, we report results from multiple interviews and a survey study (n=32) conducted with practicing oncologists. Based on these results, we compiled a set of design criteria for precision oncology reports and developed a prototype report design using these criteria, along with feedback from oncologists.

## 1 INTRODUCTION

Recent advances in next-generation sequencing (NGS) technologies [1] are improving our understanding of the role of inter-individual variability in the genomic lesions that drive disease predisposition or progression. Coupled with their low cost, genomic information from these technologies is increasingly being utilized in the clinic [2]. In fact, past several years have witnessed an emerging shift towards precision medicine approaches, which refer to treatments that are precisely tailored for each individual patient based on his or her unique genomic or epigenomic characteristics and lifestyle [3]–[5]. Since the mutational landscape of tumor cells plays an important role in the diagnosis, prognosis and treatment of multiple cancers, precision medicine approaches are especially well-suited for oncology practice. Under the remit of the Clinical Laboratory Improvement Amendments (CLIA) of 1988, genetic tests that are deemed to be analytically valid (e.g. accuracy and precision of gene mutation detection) by Centers for Medicare and Medicaid Services (CMS) can be adopted without any review of data regarding their clinical utility [6], and the utilization of test results to guide therapy resides entirely within the discretion of the treating physician. Consequently, gene panels that typically involve the targeted sequencing of 25 to 400 known cancer related genes are offered by both commercial vendors such as Caris (Figure 1—www.carislifesciences.com), Foundation Medicine (www.foundationmedicine.com), GeneDx (www.genedx.com) [7], [8] and medical centers. Among the commercial vendors, the most frequently used panel is that from FoundationOne (Figure 1), which has produced more than 22,000 somatic tests in 2014 [9].

---

\* email: selim.kalayci@mssm.edu
† email: cagatay@csail.mit.edu
‡ email: zeynep.gumus@mssm.edu

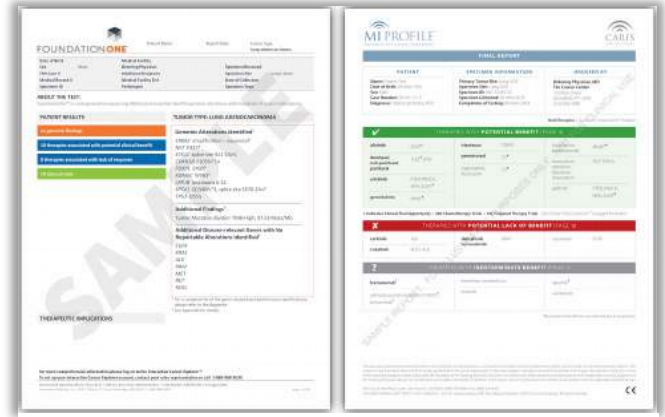

Figure 1: First page of sample reports from Foundation Medicine (left); Caris (right). See Supplementary Figure S1 and S2 respectively for full sample reports.

Results obtained from gene panel tests are typically delivered to clinicians as static print reports. So far there are no established standards, specifications or FDA guidelines. However, we can categorize the content commonly found in today's existing vendor reports into 5 main sections: patient/test identification, genetic alterations (mutations), therapeutic implications, test-related information, and appendix containing references and disclaimers (Table 1). Some vendor reports also provide one or more of the following: prognostic information, toxicity information, and pathologist interpretation of results. However, these are not necessarily included across different vendor reports. Since somatic genetic alterations form the basis of these tests, details provided on specific mutations drive most of the technical information presented in the reports.

The contents of the gene panel tests are critical as they alter treatment decisions in the clinic (i. e. hormone receptor targeting of hormone receptor positive breast tumors). However, challenges remain in communicating and interpreting these results. Several studies that investigate and report physicians' attitudes toward genetic testing and related concerns in the clinical care and practice of oncology suggest issues related with interpretation. In one of the earliest studies [10], only 29% of physicians reported that they felt qualified to provide genetic counselling to their patients, and nearly 75% of physicians thought that clear guidelines are not available for managing patients with positive test results. Similarly, more than 89% of physicians reported a need for guidelines. In a more recent study, Miller et al [11] conducted structured interviews with 17 physicians about genomic testing in oncology, and the results indicated a need for decision guidelines and education to assist physicians.

Table 1. Contents of a Typical Precision Oncology Report

| Main Section | Sub-Sections | Details |
|---|---|---|
| Patient/Test Identification | - Patient Information<br>- Specimen Information<br>- Physician/Pathologist Information | - name, gender, id, etc.<br>- specimen site, type, collection date, etc.<br>- name, contact information, etc. |
| Genetic Alterations (Mutations) | - Detected gene alterations<br>- Not detected gene alterations | - gene, type of mutation, interpretation<br>- disease-relevant gene |
| Therapeutic Implications | - FDA Approved Therapies in patient's tumor type<br>- FDA Approved Therapies in different tumor type<br>- Clinical Trials | - drug name, interpretation<br>- drug name, interpretation<br>- title, phase, location, id, etc. |
| Test-related Information | - Gene List<br>- Methodology | - list of all genes assayed in the test<br>- specific technologies and procedures used to generate the test results |
| Appendix | - References<br>- Disclaimers | - scientific works referred throughout the report<br>- FDA status, legal/technical liabilities, etc. |

In a larger study, Gray et al. [12] surveyed 160 clinically active oncologists at an academic cancer center to assess their current use of somatic testing, and their genomic confidence: 22% reported low confidence in their genomic knowledge. This study underscored a need for both evidence-based guidelines and enhanced genomics education efforts for physicians. In fact, genomic confidence was one of the strongest predictors of physicians' attitudes about and anticipated use of genetic testing. As is the case for the content, there are also no standards for the design and presentation of information in precision oncology reports. The layout, organization, design and representation of information at all levels vary widely across different vendors. In some instances, the results may indicate mutations in a large number of genes across multiple pathways, generating massive amounts of information [13]. For busy practicing oncologists, handling and interpreting such detailed and vast volumes of information presented in a nonstandard and impractical manner becomes a significant challenge. Thus, employing practical and effective reports is a crucial factor in the clinical adoption and utilization of precision oncology tests by oncologists.

To our knowledge, there are arguably no vendor-neutral/non-commercial research efforts aimed to improve and evaluate the effectiveness of precision oncology reports. On the other hand, multiple studies have focused on radiology reports. In [14], clinicians were asked to rank a variety of hypothetical radiology reports in order of preference. Reports presented in tables as opposed to traditional prose format were preferred and clinicians also preferred more detailed reports that included a clinical comment by the radiologist, for both normal and abnormal results. In a similar study [15], referring clinicians and radiologists indicated that structured reports provided better content and greater clarity than reports in simple text. In a study [16] that involved general practitioners (GP) who were asked to rank preferences for ultrasound reports with differing formats and levels of detail, the results again clearly suggested that GPs preferred detailed reports in a tabulated format.

Here, we investigated the challenges oncologists face in interpreting precision oncology reports and provided visualization solutions to some of these issues in a two-stage process, where we first identified design issues and requirements and then developed a design prototype. At every step of this process, we solicited feedback from the target audience, practicing oncologists.

## 2 DESIGN ISSUES AND EXPECTATIONS

In the first phase of our study, we aimed to understand the current status, needs and expectations of oncologists in terms of precision oncology report usage in their clinical practice. For this purpose, we conducted informal interviews with several oncologists at

Mount Sinai Health System (New York, USA) regarding their current experience, and expectations from the design of precision oncology reports. Based on these initial round of feedbacks, we prepared a draft survey questionnaire to get input from a wider community of oncologists. We revised the draft survey based on feedback from the same group to improve the relevance of the survey questions to the tasks of oncologists. After several rounds of revisions, we finalized the survey questionnaire.

### 2.1 Survey and Participants

We conducted the survey at a Hematology/Medical Oncology Grand Rounds meeting at Mount Sinai Health System in April 2016. After discussing the goals and scope of the project, the survey questionnaire was provided to all oncologists present at the meeting to fill voluntarily. The complete survey questionnaire can be found in Supplementary Table 1. In total, 32 oncologists responded to the survey.

Table 2 summarizes the demographics of the survey participants. The participants were almost exclusively made up of medical oncologists, majority of which were male (71%). Post-medical school years of experience exhibited a wide-range and were almost evenly distributed. The most commonly listed clinical foci among participating oncologists were blood cancer (42%), hematology (23%), and breast cancer (19%), where 31% of oncologists declared themselves to have multiple clinical foci.

Table 2. Demographics of Oncologist Survey (N = 32)

| Demography | Respondents |
|---|---|
| Gender | (n = 31) |
| Male<br>Female | 22<br>9 |
| Years since medical school | (n = 32) |
| 0 - 5<br>6 - 10<br>11 - 15<br>16 - 20<br>21 - 25<br>26 - 30<br>31 - 35<br>36 - 40 | 5<br>8<br>2<br>2<br>3<br>4<br>4<br>4 |
| Type of Physician (multiple answers allowed) | (n = 31) |
| Surgical Oncologist<br>Medical Oncologist<br>Radiation Oncologist | 1<br>31<br>0 |

| Design Goal | Explanation |
|---|---|
| **DG-1:** Present Genetic alteration (Biomarker) information in the best possible way | This is the most important information for oncologists (Fig. S3.4), and also the most time-consuming component during interpretation (Fig. S3.5). Thus, relevant information should be presented in a concise, non-confusing manner. Also, the whole component (especially the actionable items) should be made prominent in the overall layout of the report. |
| **DG-2:** Distinguish clinically-relevant information from non-relevant technical information | Verbosity of technical information is the most troublesome aspect of reading these reports (Fig. S3.6). Oncologists expressed that they end up sifting through a lot of data to access information that is clinically important. Thus, the overall layout should keep essential (clinically relevant) information together before non-essential (e.g. standard disclaimers, technology-related explanation, etc.) information. |
| **DG-3:** Actionable summary results should be easily discoverable | Ease of locating actionable summary and critical information is among the most troublesome aspects of reports (Fig. S3.6). Actionable summary result information should be kept together, centralized, and emphasized with design elements. |
| **DG-4:** Presentation of information should be natural and easy-to-follow for oncologists | Organization and presentation of technical information should be in accordance with the customs and experiences of oncologists. Also, design elements used with an aim to help structure/organize information should not interfere with the interpretation activities of oncologists. |
| **DG-5:** As much as possible, the design should be consistent/compatible with existing vendor reports | Unless there is good reasoning not to do so, commonly used and well-performing design aspects of existing vendor reports can be reused. This, to a certain degree, helps with the desired standardization expectations (Fig. S3.8) of oncologists. |
| **DG-6:** When applicable, employ proper visual summary/representation of data | To address the verbosity aspect of technical information in existing reports (Fig. S3.6) and oncologists' expectations of summarized data (Fig. S3.8), proper visual design elements should be incorporated. Utilization of these visual elements should not be in contradiction with the objectives of previous design goals. |

## 2.2 Results

Here, we provide a summary of our survey results. A complete analysis can be found in Supplementary Figure S3.

*Genetic alterations are the most important information.* Oncologists who participated in the survey stated that the most important information they look for in a tumor profiling report is genetic alterations (84%, Figure S3.4.). Therapy information and clinical trial information were also selected by 69% and 66% respectively (multiple selections were allowed).

*Finding actionable genetic alterations is the most time-consuming task.* By a wide margin, the most time-consuming component of interpreting such a report was stated as finding actionable genetic alteration information (57%, Figure S3.5) followed by clinical trials information (29%) and information about therapies (21%).

*Verbosity and clutter limit the effectiveness of the reports.* In general, oncologists stated that they preferred more concise reports; and that the most troublesome aspect of reading these reports was the verbosity of technical information (59%, Figure S3.6). Ease of locating actionable summary (45%) and ease of locating critical information (34%) were also identified as issues, which to a certain degree supports our previous observation.

*Layout and visual encoding are the most critical elements of a vendor report.* By a wide margin, oncologists selected the layout and presentation of information (61%, Figure S3.7) as the most problematic feature of a vendor report that makes it harder to interpret its content (next: more technical details, 32%), emphasizing the importance of our overall study. Furthermore, none of the participants stated that they were satisfied with current reports (Figure S3.8) when asked regarding the effectiveness of the reports. Furthermore, 58% of oncologists suggested standardizing the design of vendor reports. Finally, 50% suggested adding visualizations to summarize various data and to improve the effectiveness of oncology reports. Even though the percentage of those suggesting visual summarization of data may not look substantially high, it is quite noteworthy to observe such a demand from a diverse demography of clinicians who are highly accustomed to dealing with extensive textual information.

## 3 REPORT DESIGN

Based on our findings from the survey study, and additional more in-depth feedback from interviews with several oncologists, we identified six main criteria for precision oncology report design, as listed in Table 3.

### 3.1 Design Process

Conforming to the design goals in Table 3, we designed several report prototypes in an iterative process. Briefly, we used different visual encoding and styling (e.g. shapes, colors, scaling, etc.) options to present information in a concise and clear format. In addition, we utilized multiple design elements (e.g. bold/larger font, shading, etc.) to highlight important information. Furthermore, we employed contrasting design elements (e.g. varying colors and shapes, etc.) to arrange the large-volume of information in an easily distinguishable manner. At each iteration, we had in-depth consultations with two oncologists and revised the designs accordingly. Among multiple prototypes that came out of this process, the one with the highest consensus is shown in Figure 2. Accordingly, the rest of the discussions regarding the report prototype (RP) refer to this specific design.

### 3.2 Design Choices

The main choices we made in the RP are summarized as follows. First, due to the conventional usage and expectations of clinicians (DG-4), the test/patient identification information was kept at the top of the page. We then designed a Results Summary section at the top left-half of the page to address DG-3. In this section, we used a table format and incorporated basic visual elements (DG-6) to summarize the information. To address DG-1, we designed a Genomic Alterations section at the top right-half of the page. This section lists all the detected genetic alterations as well as disease-relevant alterations that were not detected. Associated therapy options are also displayed using the same visual elements, to help keep the consistency of notation and information across different sections (DG-4). Therapeutic Implications follow next, taking the whole page width, because of the type and amount of information it embeds. This section closely resembles the same section in the design of FoundationOne reports (DG-5), with the addition of color-encoded display of each row depending on whether the genomic alteration is clinically actionable. Finally, overall layout and organization of information displayed in this first page of RP follows from DG-2.

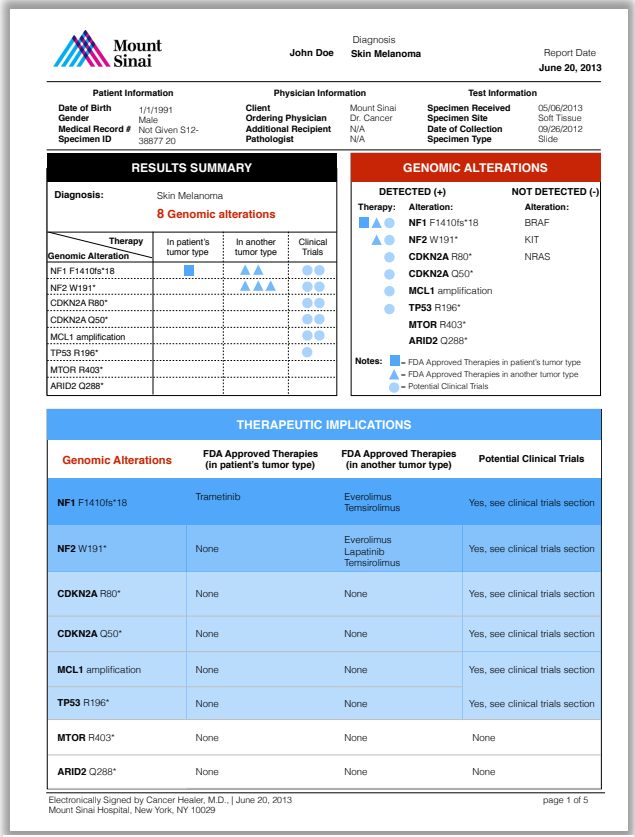

Figure 2: Our oncology report design informed by the design guidelines.

## 4 DISCUSSION AND CONCLUSION

In this study, we introduced a set of design guidelines for improving the effectiveness of precision oncology reports. To derive these guidelines, we conducted both one-on-one interviews and a systemic survey with clinical oncologists. We present here a new oncology report design informed by the derived guidelines.

How effective is our new report design? Our future work will assess the comparative merits and limitations of our design with respect to frequently used reports through a quantitative user study. Currently, like many medical reports, oncology reports are generated for clinicians. Another avenue of future research is to improve the effectiveness of these reports for patients, enabling their more informed involvement in the decision making process. Furthermore, while oncology reports are invariably viewed as static presentations in current clinical practice, we expect their evolution into interactive mediums. Therefore, future investigations on the effective design of interactive visual reports are warranted. In the meantime, our work here provides insights and visual improvements for the increased effectiveness of static reports as a reference for future studies.

## ACKNOWLEDGEMENTS

SK and ZHG gratefully acknowledge funding from NY State grant through Rensselaer CATS as well as start up funds to ZHG from Icahn School of Medicine at Mount Sinai. Authors gratefully acknowledge expert feedback from domain practitioners including Marshall Posner and MassiveBio (New York, USA).

# SUPPLEMENTARY FIGURE S1

# FOUNDATION**ONE**®

| Patient Name | Report Date | Tumor Type |
| | | **Lung adenocarcinoma** |

| Date of Birth | | Medical Facility | | Specimen Received | |
| Sex | Male | Ordering Physician | | Specimen Site | Lymph Node |
| FMI Case # | | Additional Recipient | | Date of Collection | |
| Medical Record # | | Medical Facility ID # | | Specimen Type | |
| Specimen ID | | Pathologist | | | |

## ABOUT THE TEST:
FoundationOne™ is a next-generation sequencing (NGS) based assay that identifies genomic alterations within hundreds of cancer-related genes.

## PATIENT RESULTS

**11 genomic findings**

**10 therapies associated with potential clinical benefit**

**0 therapies associated with lack of response**

**19 clinical trials**

## TUMOR TYPE: LUNG ADENOCARCINOMA

### Genomic Alterations Identified†

*ERBB2* amplification – equivocal#

*NF2* E427*

*STK11* splice site 921-1G>C

*CDKN1B* E105fs*14

*FOXP1* E490*

*KDM5C* W983*

*LRP1B* loss exons 6-14

*SPTA1* Q1346fs*3, splice site 3570-2A>T

*TP53* I255S

### Additional Findings†

*Tumor Mutation Burden* TMB-High; 37.53 Muts/Mb

### Additional Disease-relevant Genes with No Reportable Alterations Identified†

*EGFR*

*KRAS*

*ALK*

*BRAF*

*MET*

*RET*

*ROS1*

† For a complete list of the genes assayed and performance specifications, please refer to the Appendix

# See Appendix for details

## THERAPEUTIC IMPLICATIONS

**For more comprehensive information please log on to the Interactive Cancer Explorer™**
**To set up your Interactive Cancer Explorer account, contact your sales representative or call 1-888-988-3639.**

| Genomic Findings Detected | FDA-Approved Therapies (in patient's tumor type) | FDA-Approved Therapies (in another tumor type) | Potential Clinical Trials |
|---|---|---|---|
| *ERBB2* amplification - equivocal | Afatinib | Ado-trastuzumab emtansine Lapatinib Pertuzumab Trastuzumab | Yes, see clinical trials section |
| *Tumor Mutation Burden* TMB-High; 37.53 Muts/Mb | Nivolumab Pembrolizumab | Atezolizumab | Yes, see clinical trials section |
| *NF2* E427* | None | Everolimus Temsirolimus | Yes, see clinical trials section |
| *STK11* splice site 921-1G>C | None | Everolimus Temsirolimus | Yes, see clinical trials section |
| *CDKN1B* E105fs*14 | None | None | None |
| *FOXP1* E490* | None | None | None |
| *KDM5C* W983* | None | None | None |
| *LRP1B* loss exons 6-14 | None | None | None |
| *SPTA1* Q1346fs*3, splice site 3570-2A>T | None | None | None |
| *TP53* I255S | None | None | None |

Note: Genomic alterations detected may be associated with activity of certain FDA-approved drugs; however, the agents listed in this report may have little or no evidence in the patient's tumor type. Neither the therapeutic agents nor the trials identified are ranked in order of potential or predicted efficacy for this patient, nor are they ranked in order of level of evidence for this patient's tumor type.

**For more comprehensive information please log on to the Interactive Cancer Explorer™**
**To set up your Interactive Cancer Explorer account, contact your sales representative or call 1-888-988-3639.**

## GENOMIC ALTERATIONS

| GENE<br>ALTERATION | INTERPRETATION |
| --- | --- |

### ● *ERBB2*
amplification - equivocal[1]

**Gene and Alteration:** ERBB2 (also known as HER2) encodes a receptor tyrosine kinase which is in the same family as EGFR. Amplification or overexpression of ERBB2 can lead to excessive proliferation and tumor formation[1].

**Frequency and Prognosis:** In the TCGA datasets, ERBB2 amplification or mutation was observed in 6% of lung adenocarcinoma cases[2]. HER2 overexpression has been documented in 11-32% of non-small cell lung cancers (NSCLC), and is higher in lung adenocarcinomas (38%) than in squamous cell (16%) and large cell (17.9%) tumors[3,4]. A tendency toward shorter survival has been observed in patients with NSCLC harboring ERBB2 amplification and strong HER2 protein expression[5].

**Potential Treatment Strategies:** Based on extensive clinical evidence, ERBB2 amplification or activating mutation may predict sensitivity to therapies targeting HER2, including antibodies such as trastuzumab[6,7,8,9,10,11], pertuzumab in combination with trastuzumab[8,12,13], and ado-trastuzumab emtansine (T-DM1)[14], as well as dual EGFR/HER2 kinase inhibitors such as lapatinib[15,16,17,18], afatinib[11,19,20,21,22], neratinib[23,24], and dacomitinib[25]. In patients with breast cancer, concurrent PIK3CA or PTEN alterations that activate the PI3K pathway have been associated with resistance to therapies that target HER2, including trastuzumab and lapatinib[26,27,28,29,30]. However, other studies have reported conflicting results, with one study suggesting that neither PIK3CA nor PTEN alteration is associated with trastuzumab resistance[31], and another study reporting a correlation between PIK3CA mutation and increased clinical response to the combination of letrozole and lapatinib[32]. Clinical trials of agents aimed at preventing or overcoming resistance to anti-HER2 therapies are under way, including agents targeting the PI3K-AKT pathway or HSP90[33,34].

### ● *Tumor Mutation Burden*
TMB-High; 37.53 Muts/Mb

**Gene and Alteration:** Tumor mutation burden (TMB, also known as mutation load) is a measure of the number of somatic protein-coding base substitution and insertion/deletion mutations occurring in a tumor specimen. TMB is affected by a variety of causes, including exposure to mutagens such as ultraviolet light in melanoma[35,36] and cigarette smoke in lung cancer[37,38], mutations in the proofreading domains of DNA polymerases encoded by the POLE and POLD1 genes[39,40,41,42,43], and microsatellite instability (MSI)[39,42,43]. The tumor seen here harbors a high TMB. This type of mutation load has been shown to be associated with sensitivity to immune checkpoint inhibitors, including anti-CTLA-4 therapy in melanoma[44], anti-PD-L1 therapy in urothelial carcinoma[45], and anti-PD-1 therapy in non-small cell lung cancer and colorectal cancer[38,46], potentially due to expression of immune-reactive neoantigens in these tumors[38].

**Frequency and Prognosis:** High TMB has been reported in 8-13% of non-small cell lung carcinomas (NSCLCs), including 8.2-9.6% of adenocarcinomas and 8.5% of squamous cell carcinomas (SCCs) (Spigel et al., 2016; ASCO Abstract 9017, Jiang et al., 2016; ASCO Abstract e23128). High-TMB NSCLC rarely harbors known driver mutations (1% each with EGFR, ALK, ROS1, or MET), with the exception of BRAF (10.3%) or KRAS (9.4%) mutation (Spigel et al., 2016; ASCO Abstract 9017). Higher mutational load was reported to be associated with later stage NSCLC in a study of 48 African-American patients (Schwartz et al., 2016; ASCO Abstract 8533). Although some studies have reported a lack of association between smoking and mutational burden in NSCLC (Schwartz et al., 2016; ASCO Abstract 8533)[47,48], several other large studies did find a strong association with increased TMB[49,50,51,52]. A large study of Chinese patients with lung adenocarcinoma reported a shorter median overall survival (OS) for tumors with a higher number of mutations in a limited gene set compared with lower mutation number (48.4 vs. 61.0 months)[47].

**For more comprehensive information please log on to the Interactive Cancer Explorer™**
**To set up your Interactive Cancer Explorer account, contact your sales representative or call 1-888-988-3639.**

| GENE<br>ALTERATION | INTERPRETATION |
|---|---|

**Potential Treatment Strategies:** On the basis of emerging clinical evidence, increased TMB may be associated with greater sensitivity to immunotherapeutic agents, including anti-CTLA-4[44], anti-PD-L1[45], and anti-PD-1 therapies[38,46]; FDA-approved agents include ipilimumab, atezolizumab, pembrolizumab, and nivolumab. In multiple solid tumor types, higher mutational burden has corresponded with response and improved prognosis. Pembrolizumab improved progression-free survival (14.5 vs. 3.4-3.7 months) in patients with non-small cell lung cancer (NSCLC) and higher mutational load (greater than 200 nonsynonymous mutations; hazard ratio = 0.19)[38]. In studies of patients with either NSCLC or colorectal cancer (CRC), patients whose tumors harbor elevated mutational burden reported higher overall response rates to pembrolizumab[38,46]. Anti-PD-1 therapies have achieved clinical benefit for certain patients with high mutational burden, including 3 patients with endometrial adenocarcinoma who reported sustained partial responses following treatment with pembrolizumab[53] or nivolumab[54] and two patients with biallelic mismatch repair deficiency (bMMRD)-associated ultrahypermutant glioblastoma who experienced clinically and radiologically significant responses to nivolumab[55]. In patients with melanoma, mutational load was associated with long-term clinical benefit from ipilimumab[44,56] and anti-PD-1 treatment (Johnson et al., 2016; ASCO Abstract 105). For patients with metastatic urothelial carcinoma, those who responded to atezolizumab treatment had a significantly increased mutational load [12.4 mutations (mut) per megabase (Mb)] compared to nonresponders (6.4 mut/Mb)[45].

---

● **NF2**
E427*

**Gene and Alteration:** Merlin, encoded by NF2, coordinates cell contact with growth signals; the inactivation of Merlin disrupts this mechanism and can lead to unrestrained growth despite cell contact[57]. NF2 alterations that disrupt the FERM domain (amino acids 22-311), including in-frame deletions that disrupt the Paxillin-binding region (aa 50-70) of the FERM domain[58], and/or the C-terminal region (amino acids 506-547), such as observed here, are predicted to be inactivating[58,59,60,61,62,63,64]. Heterozygous germline NF2 loss or inactivation is associated with neurofibromatosis type 2 syndrome, which results in the development of vestibular schwannomas, meningiomas, ependymomas, and ocular disturbances[65,66,67]. Prevalence for this disorder in the general population is estimated to be 1:25,000[67]. In the appropriate clinical context, germline testing of NF2 is recommended.

**Frequency and Prognosis:** NF2 mutation or homozygous loss is not common in lung non-small cell lung cancer (NSCLC) and has been reported in ~1% of squamous cell carcinoma and adenocarcinoma samples analyzed in the TCGA datasets[2,68]. In one study, NF2 mutation has been reported in just 1/45 lung cancer cases[69].

**Potential Treatment Strategies:** On the basis of strong clinical evidence from multiple case reports[70,71,72,73] as well as extensive preclinical evidence[74,75], NF2 inactivation may predict sensitivity to mTOR inhibitors, including approved agents everolimus and temsirolimus. Loss or inactivation of NF2 may also predict sensitivity to FAK inhibitors, based on clinical data in mesothelioma (Soria et al., 2012; ENA Abstract 610) and strong preclinical data[76,77]. Limited preclinical and clinical evidence in vestibular schwannoma suggests possible sensitivity of NF2-deficient tumors to the pan-ERBB inhibitor lapatinib[78,79]. Similarly, on the basis of limited clinical (Subbiah et al., 2011; ASCO Abstract 2100) and preclinical[80,81,82] evidence, NF2 inactivation may predict sensitivity to MEK inhibitors, such as approved agents trametinib and cobimetinib. These and other relevant compounds are being investigated in clinical trials. A Phase 1b trial of a combination of the mTOR inhibitor everolimus and the MEK inhibitor trametinib in patients with solid tumors reported frequent adverse events and was unable to identify a recommended Phase 2 dose and schedule for the combination[83].

---

**For more comprehensive information please log on to the Interactive Cancer Explorer™**
**To set up your Interactive Cancer Explorer account, contact your sales representative or call 1-888-988-3639.**

| GENE<br>ALTERATION | INTERPRETATION |
|---|---|

● **STK11**
splice site 921-1G>C

**Gene and Alteration:** The serine/threonine kinase STK11 (also called LKB1) activates AMPK and negatively regulates the mTOR pathway in response to changes in cellular energy levels[84]. LKB1 acts as a tumor suppressor in cancer, as loss of function promotes proliferation and tumorigenesis[85,86]. Functional disruption of the STK11 kinase domain (amino acids 49-309) or STRAD binding domain (amino acids 320-343) through mutation or loss, as observed here, is predicted to be inactivating[87,88,89,90,91,92,93,94,95,96,97]. Germline mutations in STK11 underlie Peutz-Jeghers syndrome (PJS), a rare autosomal dominant disorder associated with a predisposition for tumor formation[98]. This disorder has an estimated frequency between 1:29,000 and 1:120,000, although reported rates in the literature vary greatly. Although gastrointestinal tumors are the most common malignancies associated with PJS, patients also exhibit an 18-fold increased risk of developing other epithelial cancers[98,99,100], and individuals with this syndrome have a 30-50% risk of developing breast cancer[98,100]. Given the association with PJS, in the appropriate clinical context testing for the presence of germline mutations in STK11 is recommended.

**Frequency and Prognosis:** Several clinical studies have found STK11 mutation to be common in non-small cell lung cancer (NSCLC) (15-35%), with alterations more prevalent in lung adenocarcinomas (13-34%) than in lung squamous cell carcinoma (2-19%)[51,68,101,102,103,104,105]. STK11 mutations in NSCLC often co-occur with activating KRAS mutations[104,105]. In transgenic mouse models, animals expressing mutant KRAS developed lung adenocarcinomas, whereas the KRAS-mutant/LKB1-deficient mice developed an expanded histological spectrum of tumors that included large cell and squamous cell carcinomas[102]. Decreased expression of LKB1 correlates with poor prognosis and/or higher histological grade in patients with some cancer types, although prognosis in patients with NSCLC is not known[106,107].

**Potential Treatment Strategies:** Increased mTOR signaling is present in LKB1-deficient tumors, suggesting therapies targeting mTOR may be relevant for tumors with STK11 alterations[84,102,108,109,110]. The mTOR inhibitors everolimus and temsirolimus are FDA approved for the treatment of other tumor types, and are being investigated in clinical trials for several indications[111,112,113,114]. A PJS patient with pancreatic cancer and an STK11 mutation experienced a partial response to the mTOR inhibitor everolimus[115]. Loss of STK11 also leads to activation of the downstream kinase SRC, suggesting that inhibitors such as dasatinib or bosutinib may be relevant for the treatment of LKB1-deficient tumors[85].

● **CDKN1B**
E105fs*14

**Gene and Alteration:** CDKN1B encodes the cyclin-dependent kinase inhibitor p27, which controls cell cycle progression through G1 phase by binding to prevent action of cyclin E/CDK2 and cyclin D/CDK4 protein complexes. Removal of this inhibition is required for cellular transition from quiescence to a proliferative state. There is some evidence that germline variants in CDKN1B are associated with increased risk for several tumor types, including prostate[116], endometrial[117], and colorectal cancers[118].

**Frequency and Prognosis:** Somatic inactivating mutations in CDKN1B have been documented in fewer than 1% of tumors (COSMIC, 2016). A survey of 350 breast cancers found somatic mutations in CDKN1B in approximately 1% of cases[119]. Mutations in p27 have been associated with multiple endocrine neoplasia syndrome, and truncating alterations have been shown to disrupt normal subcellular localization of p27 due to the loss of a nuclear localization motif[120,121]. Loss of p27 expression has been described in some studies as a negative indicator of prognosis in patients with B-cell lymphomas, but the relationship between p27 levels and cell proliferation is somewhat controversial[122]. Changes in the levels of p27 have been observed in the context of multiple myeloma, and decreased levels of p27 are associated with reduced overall survival and more aggressive cancers[123,124,125]. A preclinical study showed that p27 is essential for cell cycle arrest of T-cell acute lymphoblastic leukemia (T-ALL) cells by glucocorticoid treatment[126].

**For more comprehensive information please log on to the Interactive Cancer Explorer™**
**To set up your Interactive Cancer Explorer account, contact your sales representative or call 1-888-988-3639.**

| GENE<br>ALTERATION | INTERPRETATION |
| --- | --- |

**Potential Treatment Strategies:** There are no targeted therapies available to address genomic alterations in CDKN1B.

---

### ● *FOXP1*
E490*

**Gene and Alteration:** FOXP1 encodes the protein 'forkhead box protein P1', a transcription factor previously reported as a tumor suppressor, but one which can also function as an oncogene when shorter isoforms are expressed[127,128].

**Frequency and Prognosis:** Loss of FOXP1 expression has been reported to be a frequent event in endometrial cancer[129]. FOXP1 translocations have been described in acute lymphoblastic leukemia[130,131], and deletions of the chromosomal region where FOXP1 is located have been reported in acute myeloid leukemia and myeloproliferative neoplasms[132,133]. Genomic rearrangements that disrupt the 5' regulatory region of FOXP1 have been detected and characterized in several lymphomas[134,135,136]. Such alterations have been demonstrated to result in expression of N-terminally truncated variants of FOXP1, or aberrant expression of full length FOXP1 driven by strong regulatory elements, such as IGH, as observed in the t(3;14)(p13;q32) translocation[137]. In a genome-wide association study, polymorphisms at the FOXP1 locus were found to be significantly associated with Barrett esophagus and esophageal adenocarcinoma[138]. Conflicting data have been presented on the prognostic impact of FOXP1 expression, as high expression of FOXP1 is associated with poor prognosis in patients with cutaneous large B-cell lymphomas or mucosal tissue-associated lymphoid tissue (MALT) lymphomas, but improved prognosis in patients with breast or lung cancer[134,135,139,140,141].

**Potential Treatment Strategies:** There are no approved therapies available to address alterations in FOXP1.

---

### ● *KDM5C*
W983*

**Gene and Alteration:** KDM5C encodes a histone lysine demethylase that acts, along with related histone-modifying enzymes, to control gene expression in response to developmental and environmental cues[142]. In addition to its role as a histone-modifying demethylase, KDM5C has been suggested to play a role in regulation of the SMAD3 signal transduction response to TGF-beta, a role that would be consistent with function as a tumor suppressor[143]. Germline inactivating mutations in KDM5C cause an X-linked intellectual disability syndrome also characterized by short stature and hyperreflexia[144].

**Frequency and Prognosis:** Somatic mutations of KDM5C have been observed in a number of solid tumors and the role of KDM5C inactivation has been well characterized in clear cell renal cell carcinoma (ccRCC)[145,146,147,148]. However, KDM5C amplification and overexpression has been implicated in prostate cancer where KDM5C has been associated with poor prognosis[149].

**Potential Treatment Strategies:** There are no targeted therapies available to address genomic alterations in KDM5C.

---

| GENE ALTERATION | INTERPRETATION |
|---|---|

● ***LRP1B***
loss exons 6-14

**Gene and Alteration:** LRP1B encodes the low-density lipoprotein receptor-related protein 1B, also called LRPDIT. LRP1B is subject to frequent mutation, deletion, and/or silencing in cancers, leading to the hypothesis that it behaves as a tumor suppressor. However, the mechanism of tumor suppression is unclear[50,150,151]. The LRP1B protein consists of three regions: an extracellular LDL-receptor (amino acids 25-4444), a transmembrane region (4445-4467), and a smaller cytoplasmic portion (4468-4599). Somatic mutations that lead to C-terminal truncation of LRP1B are common, and in vitro studies suggest that an intracellular domain fragment released by a gamma-secretase-like activity translocates to the nucleus where it suppresses anchorage-independent cell growth[152]. In addition, heterozygous mice that express a C-terminally truncated LRP1B missing codons 3547- 4599, which includes the transmembrane and cytoplasmic domain were reported to be viable, with no phenotype; however, mice homozygous for the mutation or the null mutation were inviable[153]. Therefore, it is possible that truncated proteins are still functional.

**Frequency and Prognosis:** LRP1B mutations have been frequently reported in many types of cancer, including 12-16% of multiple myeloma[154,155], 6% of diffuse large B-cell lymphoma, 3% of chronic lymphocytic leukemia/small cell lymphoma, 1% of acute myeloid leukemia, and 7% (2/29) of chronic myeloid leukemia cases (COSMIC, 2016). In addition, LRP1B mutations have been frequently reported in many solid tumors: 32% of melanoma, 30-39% of squamous cell lung cancer, 28-32% of lung adenocarcinoma, 26% of stomach cancer, and 6-20% of head and neck squamous cell carcinoma, bladder cancer, and colorectal cancer cases (cBioPortal, 2016). LRP1B is commonly inactivated in non- small cell lung cancer cell lines, and low expression of LRP1B mRNA was associated with poor patient outcome[156,157,158].

**Potential Treatment Strategies:** There are no therapies or clinical trials that address the loss or mutation of LRP1B in cancer. In some tumor types, such as high-grade serous cancer (HGSC), LRP1B deletion has been reported to be associated with resistance to liposomal doxorubicin[159].

● ***SPTA1***
Q1346fs*3, splice site 3570-2A>T

**Gene and Alteration:** SPTA1 encodes the protein spectrin alpha chain 1, a component of the cytoskeleton of erythrocytes[160]. Germline mutations in SPTA1 have been associated with disorders featuring abnormally shaped erythrocytes, such as elliptocytosis and spherocytosis[161].

**Frequency and Prognosis:** SPTA1 mutations have been found in 9% of glioblastoma samples analyzed in one study[162], and mutations have been reported with high prevalence in melanoma, lung tumors, and esophageal tumors (COSMIC, cBioPortal, 2016).

**Potential Treatment Strategies:** There are no therapies available to directly address genomic alterations in SPTA1 in cancer.

● ***TP53***
I255S

**Gene and Alteration:** Functional loss of the tumor suppressor p53, which is encoded by the TP53 gene, is common in aggressive advanced cancers[163]. Mutations affecting the DNA binding domain (aa 100-292), the tetramerization domain (aa 325-356), or the C-terminal regulatory domain (aa 356-393), such as observed here, are thought to disrupt the transactivation of p53-dependent genes and are predicted to promote tumorigenesis[164,165,166,167]. Germline mutations in TP53 are associated with the very rare disorder Li-Fraumeni syndrome and the early onset of many cancers[168,169,170,171,172,173]. Estimates for the prevalence of germline TP53 mutations in the general population range from 1:5,000[174] to 1:20,000[173], and in the appropriate clinical context, germline testing of TP53 is recommended.

| GENE<br>ALTERATION | INTERPRETATION |
|---|---|

**Frequency and Prognosis:** TP53 is one of the most commonly mutated genes in lung cancer, and mutations in this gene have been reported in 43-80% of non-small cell lung cancers (NSCLCs)[2,175,176,177,178] and specifically in 45% of lung adenocarcinoma samples[179,180]. Mutations in TP53 have been associated with lymph node metastasis in patients with lung adenocarcinoma[181].

**Potential Treatment Strategies:** There are no approved therapies to address TP53 mutation or loss. However, tumors with TP53 loss of function alterations may be sensitive to the WEE1 inhibitor AZD1775[182,183,184,185], therapies that reactivate mutant p53 such as APR-246[186], or p53 gene therapy and immunotherapeutics such as SGT-53[187,188,189,190] and ALT-801 (Hajdenberg et al., 2012; ASCO Abstract e15010). Combination of AZD1775 with paclitaxel and carboplatin achieved significantly longer progression-free survival than paclitaxel and carboplatin alone in patients with TP53-mutant ovarian cancer (Oza et al., 2015; ASCO Abstract 5506). Furthermore, AZD1775 in combination with carboplatin achieved a 27% (6/22) response rate and 41% (9/22) stable disease rate in patients with TP53-mutant ovarian cancer refractory or resistant to carboplatin plus paclitaxel (Leijen et al., 2015; ASCO Abstract 2507). In a Phase 1 clinical trial, 8 of 11 evaluable patients receiving SGT-53 as a single agent exhibited stable disease[191]. Clinical trials of SGT-53 in combination with chemotherapy are underway. Additionally, the combination of a CHK1 inhibitor and irinotecan reportedly reduced tumor growth and prolonged survival in a TP53 mutant, but not TP53 wild-type, breast cancer xenotransplant mouse model[192]. Kevetrin has also been reported to activate p53 in preclinical studies and might be relevant in the context of mutant p53 (Kumar et al., 2012; AACR Abstract 2874). Clinical trials of these agents are under way for some tumor types for patients with a TP53 mutation.

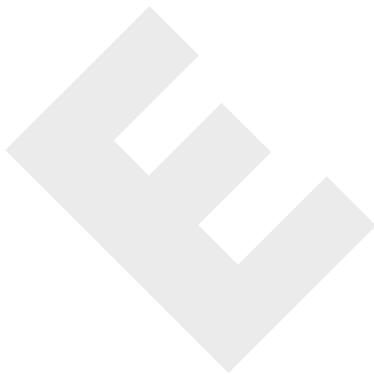

## THERAPIES

### FDA-APPROVED THERAPIES IN PATIENT TUMOR TYPE

| THERAPY | SUMMARY OF DATA IN PATIENT TUMOR TYPE |
|---|---|
| Afatinib | **Approved Indications:** Afatinib is an irreversible kinase inhibitor that targets the kinase domains of EGFR, ERBB2/HER2, and ERBB4. It is FDA approved for the treatment of metastatic non-small cell lung cancer (NSCLC) in patients with EGFR exon 19 deletions or exon 21 (L858R) missense mutations.<br><br>**Gene Association:** ERBB2 amplification or activating mutations may indicate sensitivity to afatinib on the basis of clinical evidence in various solid tumors[11,19,193].<br><br>**Supporting Data:** Phase 3 clinical trials have demonstrated that treatment with afatinib, compared to chemotherapy, leads to significantly increased progression-free survival for patients with EGFR-mutant NSCLC[194,195], and increased overall survival (OS) for patients with EGFR exon 19 alterations specifically[196]. A Phase 3 trial comparing afatinib with erlotinib as second-line therapies for advanced lung squamous cell carcinoma reported significantly higher OS (7.9 months vs. 6.8 months) and disease control rate (DCR) (51% vs. 40%) for patients treated with afatinib[197]. Phase 2/3 studies of afatinib treatment for patients with erlotinib- or gefitinib-resistant NSCLC have generally reported partial responses (PRs) of only 7-9%[22,198,199,200,201,202], and DCRs of more than 50%[22]; in particular, disease control was achieved for 2/2 patients with EGFR-amplified NSCLC[22] and 9/14 patients with T790M-positive NSCLC[202]. The T790M mutation has been implicated in reduced response to afatinib[201,203,204], with a secondary T790M mutation reported in 48% (20/42) of patients with afatinib-resistant lung adenocarcinoma[203]. The combination of afatinib with cetuximab resulted in a higher response rate (29%) for patients with erlotinib- or gefitinib-resistant disease[205], including T790M-positive cases[205,206], although adverse reactions may be a concern with this combination[207]. Upon progression on afatinib, further benefit has been reported from combination treatment with afatinib and paclitaxel[208]. |
| Nivolumab | **Approved Indications:** Nivolumab is a monoclonal antibody that binds to the PD-1 receptor and blocks its interaction with PD-L1 and PD-L2, thereby reducing inhibition of the antitumor immune response. It is FDA approved to treat unresectable or metastatic melanoma as both a single agent and in combination with the immunotherapy ipilimumab. Nivolumab is also approved to treat non- small cell lung cancer (NSCLC) following disease progression on prior treatments, advanced renal cell carcinoma following antiangiogenic therapy, and classical Hodgkin lymphoma (cHL) that has relapsed or progressed after autologous hematopoietic stem cell transplantation (HSCT) and post-transplantation brentuximab vedotin.<br><br>**Gene Association:** On the basis of emerging clinical data in patients with non-small cell lung cancer (Spigel et al., 2016; ASCO Abstract 9017)[38], colorectal cancer[46], or melanoma (Johnson et al., 2016; ASCO Abstract 105) and case reports in endometrial cancer[53,54] and glioblastoma[55], high tumor mutation burden (TMB) may predict sensitivity to anti-PD-1 therapies such as nivolumab. |

**Supporting Data:** Studies investigating the use of nivolumab as first-line treatment for patients with non-small cell lung carcinoma (NSCLC) reported an objective response rate (ORR) of 23% (12/53), median overall survival (OS) of 19.4 months, and 1-year OS rate of 73% with monotherapy[209]; combinations with platinum-based doublet chemotherapy (gemcitabine/cisplatin, pemetrexed/cisplatin, and paclitaxel/carboplatin) resulted in ORRs of 33-47%, 1-year OS rates of 50-87%, and 2-year OS rates of 25-62%[210]. In patients with platinum-refractory non-squamous NSCLC, nivolumab improved median OS (12.2 vs. 9.4 months) and the ORR (19% vs. 12%) compared with docetaxel; PD-L1 expression was associated with benefit from nivolumab in this study (OS hazard ratios of 0.40-0.59)[211]. In patients with previously treated squamous NSCLC, nivolumab resulted in longer median OS (9.2 vs. 6.0 months) and higher ORR (20% vs. 9%) than docetaxel, and PD-L1 expression was neither prognostic nor predictive of nivolumab efficacy[212,213]. Real-world studies of nivolumab for the treatment of NSCLC reported clinical benefit for 35-36% of patients (Crino et al., 2016; ASCO Abstract 3067, Corny et al., 2016; ASCO Abstract e20633). A Phase 1 study of nivolumab (3 mg/kg) plus ipilimumab (1 mg/kg, every 6 or 12 weeks) as first-line treatment for advanced NSCLC resulted in ORRs of 31-39% and median PFS of 8.0-8.3 months (Hellmann et al., 2016; ASCO Abstract 3001). Nivolumab in combination with erlotinib for the treatment of chemotherapy-naive EGFR-mutant NSCLC achieved an ORR of 19%; additionally, 15% (3/20) partial responses (PR) and 45% (9/20) stable diseases were reported in cases with acquired erlotinib resistance (Rizvi et al., 2014; ASCO Abstract 8022). Nivolumab has shown intracranial activity, with disease control in the brain for 33% of patients (Goldman et al., 2016; ASCO Abstract 9038)[214]. A small study of 3 patients with resectable NSCLC reported 1 complete response and 1 PR with nivolumab as neoadjuvant therapy (Forde et al., 2016; ASCO Abstract e20005).

**Pembrolizumab**

**Approved Indications:** Pembrolizumab is a monoclonal antibody that binds to the PD-1 receptor and blocks its interaction with the ligands PD-L1 and PD-L2 to enhance antitumor immune responses. It is FDA approved to treat unresectable or metastatic melanoma and PD-L1-positive metastatic non- small cell lung cancer (NSCLC) refractory to prior therapy.

**Gene Association:** On the basis of emerging clinical data in patients with non-small cell lung cancer (Spigel et al., 2016; ASCO Abstract 9017)[38], colorectal cancer[46], or melanoma (Johnson et al., 2016; ASCO Abstract 105) and case reports in endometrial cancer[53,54] and glioblastoma[55], high tumor mutation burden (TMB) may predict sensitivity to anti-PD-1 therapies such as pembrolizumab.

**Supporting Data:** In a Phase 2/3 study for previously treated NSCLC with PD-L1 expression (on at least 1% of tumor cells), pembrolizumab extended median overall survival (OS) (10.4-12.7 vs 8.2 months) when compared with docetaxel[215]. A Phase 1 study of pembrolizumab in NSCLC reported an overall response rate (ORR) of 19%, and median OS of 10.6 months and 22.1 months for previously treated and treatment-naive patients, respectively (Hui et al., 2016; ASCO Abstract 9026, Garon et al., 2016; ASCO Abstract 9024)[216]. In both studies, pembrolizumab demonstrated greater efficacy in patients with PD-L1 expression on at least 50% of tumor cells, with ORRs (29-45%)[215,216], median OS (14.9-17.3 months)[215], and progression-free survival (PFS; 5.0-6.3 months)[215,216] being increased for these patient populations. In a Phase 2 study of pembrolizumab for advanced PD-L1-positive NSCLC with brain metastases, 33% (6/18) of patients experienced brain metastases responses[217]. As first-line therapy for patients with EGFR/ALK wild-type advanced NSCLC, pembrolizumab plus platinum doublet chemotherapy (carboplatin/paclitaxel, carboplatin/paclitaxel/bevacizumab, or carboplatin/pemetrexed) achieved ORRs of 52% (13/25) and 59% (29/49) for patients with any histology or with nonsquamous NSCLC, respectively (Gadgeel et al., 2016; ASCO Abstract 9016). Pembrolizumab combined with the anti-CTLA4 antibody ipilimumab for patients with recurrent advanced NSCLC and at least one previous treatment reported an ORR of 24%, stable disease rate of 40% (18/45), and median OS of 17 months (Gubens et al., 2016; ASCO Abstract 9027). A Phase 1 study of pembrolizumab in combination with the anti-4-1BB antibody utomilumab reported a partial response for 1 out of 6 cases with NSCLC (Tolcher et al., 2016; ASCO Abstract 3002).

## ADDITIONAL THERAPIES – FDA-APPROVED IN OTHER TUMOR TYPES

| THERAPY | SUMMARY OF DATA IN OTHER TUMOR TYPE |
| --- | --- |
| Ado-trastuzumab emtansine | **Approved Indications:** Ado-trastuzumab emtansine (T-DM1) is an antibody-drug conjugate that targets the protein ERBB2/HER2 on the cell surface, inhibiting HER2 signaling[218,219]; it also releases the cytotoxic therapy DM1 into cells, leading to cell death[219,220]. T-DM1 is FDA approved for the treatment of HER2-positive (HER2+) metastatic breast cancer. |
|  | **Gene Association:** ERBB2 activating mutations or amplification may predict sensitivity to T-DM1. |
|  | **Supporting Data:** A patient with non-small cell lung cancer, disease progression on two prior lines of chemotherapy, and an activating ERBB2 alteration (exon 20 insertion) experienced a rapid and durable response to T-DM1[221,222]. The vast majority of data on the therapeutic use of T-DM1 has been collected in the context of breast cancer, although clinical trials investigating T-DM1 are underway in several tumor types, primarily in HER2+ cancers. A Phase 3 trial in 602 patients with HER2+ breast cancer reported that those who received T-DM1 showed an improved progression-free survival (PFS) and a lower rate of adverse events than patients who received the physician's choice of therapy[223]. A second Phase 3 trial in 991 patients with HER2+ breast cancer reported that T-DM1 brought about significantly longer overall survival (OS) and PFS, as compared with lapatinib plus capecitabine, in patients previously treated with trastuzumab plus a taxane[14,224]. Two separate Phase 2 trials reported robust activity for single-agent T-DM1 as a treatment for HER2+ metastatic breast cancer in patients previously treated with standard HER2-directed therapies or HER2-directed therapies plus chemotherapy, with objective response rates of 34.5% and 25.9%, respectively, and PFS of 6.9 months and 4.9 months, respectively[225,226]. |
| Lapatinib | **Approved Indications:** Lapatinib is a tyrosine kinase inhibitor that targets EGFR, ERBB2/HER2, and to a lesser degree, ERBB4. It is FDA approved in combination with capecitabine or letrozole for the treatment of HER2-overexpressing (HER2+) metastatic breast cancer. |
|  | **Gene Association:** Activation or amplification of ERBB2 may predict sensitivity to lapatinib. In one study, a patient with inflammatory breast cancer and ERBB2 V777L and S310F activating mutations, but without ERBB2 amplification or protein overexpression, experienced tumor shrinkage in response to combined treatment with lapatinib and trastuzumab[18]. |
|  | **Supporting Data:** Investigations into the efficacy of lapatinib have primarily been in the context of breast cancer. In preclinical assays, lapatinib reduced cell proliferation in vitro and reduced the number and size of tumors in mouse xenograft models of EGFR- and ERBB2-amplified non-small cell lung cancer (NSCLC) cells[227]. A Phase 1 study of single-agent lapatinib included 9 unselected patients with lung cancer and reported 1 case of prolonged stable disease[228]. In a Phase 2 trial in patients with advanced or metastatic NSCLC, lapatinib monotherapy did not result in significant tumor reduction, but further investigation of lapatinib in combination with other therapies may be warranted[229]. |
| Pertuzumab | **Approved Indications:** Pertuzumab is a monoclonal antibody that interferes with the interaction between HER2 and ERBB3. It is FDA approved in combination with trastuzumab and docetaxel to treat a subset of patients with HER2-positive (HER2+) breast cancer[12]. |
|  | **Gene Association:** ERBB2 amplification or activating mutations may predict sensitivity to pertuzumab. |

**Supporting Data:** In a Phase 1 study of pertuzumab in advanced cancer, 2/19 patients reported partial responses and 6/19 patients reported stable disease after two cycles, including one patient with lung cancer[230]. In another Phase 1 study in Japanese patients with solid tumors, no responses were observed and stable disease was reported in 1 of 7 patients with NSCLC[231]. In a Phase 2 study of pertuzumab in NSCLC, no responses were observed and the progression-free survival was 6.1 weeks[232]. Phase 1 and 2 trials of pertuzumab in combination with erlotinib in NSCLC have reported a response rate of 20% (3/15, 2 of the responders had mutant EGFR)[233]; a reduction in circulating tumor cells was noted and correlated with reduction in tumor size[234]. In a Phase 2 study of pertuzumab plus erlotinib in relapsed patients with NSCLC, PET-CT imaging showed that the primary endpoint of response rate (RR) was met in 19.5% of all patients (n = 41) and in 8.7% of patients with wild-type EGFR NSCLC (n = 23); however, 68.3% (28/41) of patients showed treatment-related grade 3 (or higher) adverse events[235].

| Trastuzumab | **Approved Indications:** Trastuzumab is a monoclonal antibody that targets the protein ERBB2/HER2. It is FDA approved for the treatment of breast cancers or metastatic gastric or gastroesophageal adenocarcinomas that overexpress HER2. |

**Gene Association:** ERBB2 amplification or activating mutations may confer sensitivity to trastuzumab. Trastuzumab-involving regimens elicited significant responses in patients with NSCLC and ERBB2 exon 20 insertions (8 partial responses (PRs) and 4 stable disease vs. 1 progressive disease) and in a patient with breast cancer harboring ERBB2 V777L and S310F activating mutations[11,18]. A patient with HER2-positive parotid salivary duct carcinoma also reported a PR following treatment with trastuzumab in combination with carboplatin and docetaxel[236].

**Supporting Data:** A Phase 2 clinical trial of docetaxel with trastuzumab in non-small cell lung cancer (NSCLC) reported partial responses in 8% of patients; response did not correlate with HER2 status as assessed by immunohistochemistry[237]. Another Phase 2 study of 169 patients with NSCLC reported an objective response rate of 23% (7/30 patients) in the patients treated with a combination therapy of docetaxel and trastuzumab, and 32% (11/34) in patients treated with paclitaxel and trastuzumab[238]. HER2 expression did not impact the results of this study[238]. A patient with lung adenocarcinoma that was HER-positive by FISH and harbored an ERBB2 G776L mutation experienced a partial response on trastuzumab and paclitaxel[9]. In a retrospective analysis of patients with NSCLC harboring ERBB2 exon 20 insertion mutations, disease control was reported in 93% of patients (13/14) treated with trastuzumab in combination with chemotherapy[11].

| Atezolizumab | **Approved Indications:** Atezolizumab is a monoclonal antibody that binds to PD-L1 and blocks its interaction with PD-1 in order to enhance antitumor immune responses. It is FDA approved to treat patients with advanced urothelial carcinoma who progress during or following platinum-based chemotherapy. |

**Gene Association:** On the basis of emerging clinical data in patients with urothelial carcinoma[45], non-small cell lung cancer (Spigel et al., 2016; ASCO Abstract 9017), or melanoma (Johnson et al., 2016; ASCO Abstract 105), high tumor mutation burden (TMB) may predict sensitivity to anti-PD-L1 therapies such as atezolizumab.

**Supporting Data:** A Phase 2 study of atezolizumab for the treatment of non-small cell lung carcinoma (NSCLC) reported significantly improved median overall survival (OS; 12.6 vs. 9.7 months) and objective response duration (18.6 vs. 7.2 months) when compared with docetaxel; PD-L1 expression correlated with improved response to atezolizumab (median OS 15.1 vs. 9.7 months) (Smith et al., 2016; ASCO Abstract 9028)[239]. Patients on this study who continued on atezolizumab after experiencing progressive disease (PD) achieved responses in 11% of cases and a median OS of 11.1 months, compared with 8.3 months for patients switching to different treatment (Mazieres et al., 2016; ASCO Abstract 9032). In another study of atezolizumab in patients with NSCLC, an overall response rate (ORR) of 23% (12/53) and a median progression-free survival of 15 weeks were reported[240]. Atezolizumab achieved similar ORRs for patients with NSCLC who received no prior chemotherapy (19-29%), progressed on previous platinum therapy (17-27%), or had brain metastases or treated asymptomatic brain metastases (17%) (Besse et al., 2015; ECC Abstract 16LBA, Spigel et al., 2015; ASCO Abstract 8028).

| | |
|---|---|
| Everolimus | **Approved Indications:** Everolimus is an orally available mTOR inhibitor that is FDA approved to treat renal cell carcinoma (RCC) following antiangiogenic therapy; pancreatic neuroendocrine tumors and well-differentiated non-functional neuroendocrine tumors of the lung or gastrointestinal tract; and, in association with tuberous sclerosis complex (TSC), renal angiomyolipoma and subependymal giant cell astrocytoma. Everolimus is also approved to treat hormone receptor-positive, HER2-negative advanced breast cancer in combination with exemestane following prior therapy with letrozole or anastrozole, as well as in combination with the multikinase inhibitor lenvatinib to treat advanced RCC following prior antiangiogenic therapy. |

**Gene Association:** Preclinical data suggests that loss or inactivation of NF2 may be associated with sensitivity to rapamycin, which is similar in mechanism of action to everolimus[74,75]. Several case reports describe durable complete or partial responses of patients with NF2-mutant solid tumors to therapy regimens including everolimus or temsirolimus[70,71,72,73]. Increased mTOR signaling is present in LKB1-deficient tumors[84,102,108,110,241]; therefore, therapies targeting mTOR may be relevant for tumors with STK11 alterations[84]. Clinical responses to everolimus have been reported in patients with pancreatic cancer and STK11 alterations, with two patients exhibiting a partial response for more than 6 months (Moreira et al. 2015; ASCO Abstract 315)[115].

**Supporting Data:** A trial of everolimus as a monotherapy in non-small cell lung cancer (NSCLC) showed modest activity[242], but a Phase 2 study of everolimus in combination with docetaxel did not show any added benefit of everolimus in an unselected population (Khuri et al., 2011; ASCO Abstract e13601). A Phase 1 study evaluated the addition of everolimus to carboplatin and paclitaxel +/- bevacizumab in advanced NSCLC and found the combinations produced 1 complete response and 10 partial responses (n=52), although treatments were not well tolerated[243]. A Phase 1 study in patients with advanced NSCLC of the combination of everolimus and erlotinib reported 9 objective responses and 28 patients experiencing stable disease (n=74), but a Phase 2 study found the combination inefficacious at tolerated doses[244,245]. A trial of combination treatment with sorafenib and everolimus that included 2 patients with lung adenocarcinoma reported a partial response in one patient and stable disease in the other, with both patients experiencing progression-free survival of more than 4 months[246]. A Phase 1b trial of a combination of everolimus and the MEK inhibitor trametinib in patients with solid tumors reported frequent adverse events and the study was unable to identify a recommended Phase 2 dose and schedule for the combination[83].

| | |
|---|---|
| Temsirolimus | **Approved Indications:** Temsirolimus is an intravenous mTOR inhibitor that is FDA approved to treat advanced renal cell carcinoma. |

**Gene Association:** Preclinical data suggests that loss or inactivation of NF2 may be associated with sensitivity to rapamycin, which has a similar mechanism of action to temsirolimus[74,75]. Several case reports describe durable complete or partial responses of patients with NF2-mutant solid tumors to therapy regimens including everolimus or temsirolimus[70,71,72,73]. Increased mTOR signaling is present in LKB1-deficient tumors[84,102,108,110,241]; therefore, therapies targeting mTOR may be relevant for tumors with STK11 alterations[84].

**Supporting Data:** In a Phase 2 clinical trial in NSCLC, front-line temsirolimus monotherapy demonstrated some clinical benefit but failed to meet the trial's primary end point[247]. In a Phase 1 trial of temsirolimus and radiation in patients with NSCLC, of 8 evaluable patients, 3 exhibited a partial response, and 2 exhibited stable disease[248].

Genomic alterations detected may be associated with activity of certain approved drugs; however, the agents listed in this report may have little or no evidence in the patient's tumor type.

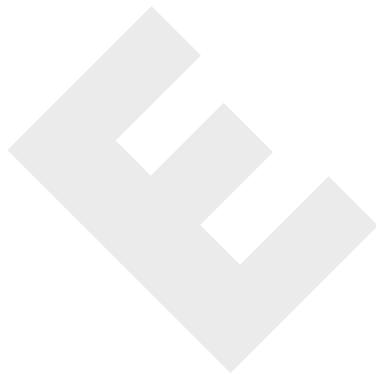

**For more comprehensive information please log on to the Interactive Cancer Explorer™**
**To set up your Interactive Cancer Explorer account, contact your sales representative or call 1-888-988-3639.**

## CLINICAL TRIALS TO CONSIDER

IMPORTANT: While every effort is made to ensure the accuracy of the information contained below, the information available in the public domain is continually updated and should be investigated by the physician or research staff. This is not meant to be a complete list of available trials. In order to conduct a more thorough search, please go to www.clinicaltrials.gov and use the search terms provided below. For more information about a specific clinical trial, type the NCT ID of the trial indicated below into the search bar.

| GENE | RATIONALE FOR POTENTIAL CLINICAL TRIALS |
|---|---|
| ***ERBB2***<br>amplification - equivocal | ERBB2 amplification or activating mutations may confer sensitivity to HER2-targeted and dual EGFR/HER2-directed therapies, and may enhance efficacy of chemotherapy or other targeted therapies, such as HSP90 inhibitors.<br><br>Examples of clinical trials that may be appropriate for this patient are listed below. These trials were identified through a search of the trial website clinicaltrials.gov using keyword terms such as "ERBB2", "HER2", "trastuzumab", "lapatinib", "pertuzumab", "ado-trastuzumab emtansine", "afatinib", "HSP90", "NSCLC", "lung", "solid tumor", and/or "advanced cancer". |

| TITLE | PHASE | TARGETS | LOCATIONS | NCT ID |
|---|---|---|---|---|
| Phase I Active Immunotherapy Trial With a Combination of Two Chimeric (Trastuzumab-like and Pertuzumab-like)Human Epidermal Growth Factor Receptor 2 (HER-2) B Cell Peptide Vaccine Emulsified in ISA 720 and Nor-MDP Adjuvant in Patients With Advanced Solid Tumors | Phase 1 | ERBB2 | Ohio | NCT01376505 |
| An Open-label, Multicenter, Multinational, Phase 2 Study Exploring the Efficacy and Safety of Neratinib Therapy in Patients With Solid Tumors With Activating HER2, HER3 or EGFR Mutations or With EGFR Gene Amplification. | Phase 2 | EGFR, ERBB2, ERBB4 | California, Florida, Massachusetts, Missouri, New Jersey, New York, Tennessee, Texas, Barcelona (Spain), Cremona (Italy), Helsinki (Finland), London (United Kingdom), Madrid (Spain), Petch Tiqwa (Israel), Rehovot (Israel), Seoul (Korea, Republic of), Torino (Italy), Valencia (Spain), Victoria (Australia) | NCT01953926 |
| Targeted Agent and Profiling Utilization Registry (TAPUR) Study | Phase 2 | ALK, Others | Michigan, North Carolina | NCT02693535 |
| My Pathway: An Open Label Phase IIa Study Evaluating Trastuzumab/Pertuzumab, Erlotinib, Vemurafenib, and Vismodegib in Patients Who Have Advanced Solid Tumors With Mutations or Gene Expression Abnormalities Predictive of Response to One of These Agents | Phase 2 | EGFR, ERBB2, BRAF, SMO | Arizona, Arkansas, California, Colorado, Florida, Georgia, Illinois, Maryland, Minnesota, New York, North Carolina, North Dakota, Ohio, Oklahoma, Oregon, Pennsylvania, South Dakota, Tennessee, Texas, Virginia, Washington | NCT02091141 |
| Phase I Trial Evaluating Safety and Tolerability of the Irreversible Epidermal Growth Factor Receptor Inhibitor Afatinib (BIBW 2992) in | Phase 1 | EGFR, ERBB2, KIT, PDGFRs, SRC, ABL | Florida | NCT01999985 |

**For more comprehensive information please log on to the Interactive Cancer Explorer™**
**To set up your Interactive Cancer Explorer account, contact your sales representative or call 1-888-988-3639.**

| Combination With the SRC Kinase Inhibitor Dasatinib for Patients With Non-small Cell Lung Cancer (NSCLC) | | | | |
|---|---|---|---|---|

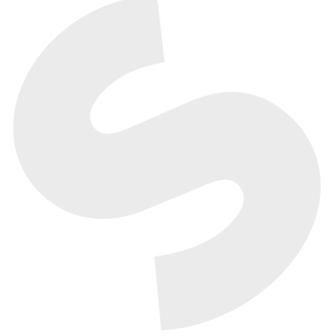

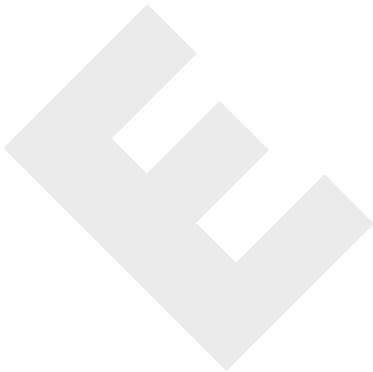

## CLINICAL TRIALS TO CONSIDER

| GENE | RATIONALE FOR POTENTIAL CLINICAL TRIALS |
| --- | --- |
| ***Tumor Mutation Burden***<br>TMB-High; 37.53 Muts/Mb | High tumor mutational burden may predict response to anti-PD-1 and anti-PD-L1 immune checkpoint inhibitors.<br><br>Examples of clinical trials that may be appropriate for this patient are listed below. These trials were identified through a search of the trial website clinicaltrials.gov using keyword terms such as "PD-L1", "B7-H1", "PD-1", "pembrolizumab", "nivolumab", "atezolizumab", "MPDL3280A", "durvalumab", "MEDI4736", "avelumab", "MSB0010718C", "BMS-936559", "CT-011", "NSCLC", "lung", "solid tumor", and/or "advanced cancer". |

| TITLE | PHASE | TARGETS | LOCATIONS | NCT ID |
| --- | --- | --- | --- | --- |
| A Phase Ib Study of the Safety and Pharmacology of Atezolizumab (Anti-PD-L1 Antibody) Administered With Bevacizumab and/or With Chemotherapy in Patients With Advanced Solid Tumors | Phase 1 | PD-1, VEGFA | Colorado, Connecticut, District of Columbia, Illinois, Massachusetts, New York, North Carolina, Tennessee | NCT01633970 |
| A Phase II Trial of Concurrent Chemoradiation With Consolidation Pembrolizumab (MK-3475) for the Treatment of Inoperable or Unresectable Stage III Non-Small Cell Lung Cancer (NSCLC): HCRN LUN14-179 | Phase 2 | PD-1 | California, Indiana, Nebraska, New Jersey | NCT02343952 |
| A Phase III, Open-Label, Randomized Study of Atezolizumab (MPDL3280A, Anti-Pd-L1 Antibody) in Combination With Carboplatin or Cisplatin + Pemetrexed Compared With Carboplatin or Cisplatin + Pemetrexed in Patients Who Are Chemotherapy-Naive and Have Stage IV Non-Squamous Non-Small Cell Lung Cancer | Phase 3 | PD-L1 | California, Connecticut, Florida, Georgia, Illinois, Indiana, Kentucky, Michigan, Nebraska, Oregon, Pennsylvania, Texas, Virginia, Washington, Wisconsin, Aichi (Japan), Alicante (Spain), Barcelona (Spain), Bunkyo-ku (Japan), Burgos (Spain), Creteil (France), Guipuzcoa (Spain), Hiroshima (Japan), Hokkaido (Japan), Hyogo (Japan), Ishikawa (Japan), Kagoshima (Japan), Kanagawa (Japan), La Coruña (Spain), Limoges (France), Malaga (Spain), Michalovce (Slovakia), Navarra (Spain), Niigata (Japan), Osaka (Japan), Piemonte (Italy), Saga (Japan), Tokyo (Japan), Yamaguchi (Japan) | NCT02657434 |
| A Randomized, Double-Blind, Phase III Study of Carboplatin-Paclitaxel/Nab-Paclitaxel Chemotherapy With or Without Pembrolizumab (MK-3475) in First Line Metastatic Squamous Non-small Cell Lung Cancer Subjects (KEYNOTE-407) | Phase 3 | PD-1 | California, Illinois, Maryland, Massachusetts, New York, North Carolina, South Carolina, Moscow (Russian Federation), North Ryde (Australia) | NCT02775435 |

**For more comprehensive information please log on to the Interactive Cancer Explorer™**
**To set up your Interactive Cancer Explorer account, contact your sales representative or call 1-888-988-3639.**

| | | | | |
|---|---|---|---|---|
| An Open-Label, Randomized Phase 3 Trial of Nivolumab, or Nivolumab Plus Ipilimumab, or Nivolumab Plus Platinum Doublet Chemotherapy Versus Platinum Doublet Chemotherapy in Subjects With Chemotherapy-Naïve Stage IV or Recurrent Non-Small Cell Lung Cancer (NSCLC) | Phase 3 | PD-1, CTLA4 | Alabama, California, Connecticut, Georgia, Kentucky, Maryland, Missouri, New Jersey, New York, North Carolina, Ohio, Pennsylvania, South Carolina, Tennessee, Texas, Utah, Washington, multiple ex-US locations | NCT02477826 |
| A Dose Frequency Optimization, Phase IIIB/IV Trial of Nivolumab 240 mg Every 2 Weeks vs Nivolumab 480 mg Every 4 Weeks in Subjects With Advanced or Metastatic Non-small Cell Lung Cancer Who Received 4 Months of Nivolumab at 3 mg/kg or 240 mg Every 2 Weeks | Phase 3 | PD-1 | Arizona, California, Colorado, Florida, Illinois, Maryland, Nebraska, New York, Ohio, Oregon, South Carolina, Tennessee, Texas, Virginia, Washington, Ontario (Canada) | NCT02713867 |
| A Phase III, Open-Label, Multicenter, Randomized Study to Investigate the Efficacy and Safety of Atezolizumab (Anti-PD-L1 Antibody) Compared With Docetaxel in Patients With Non-Small Cell Lung Cancer After Failure With Platinum-Containing Chemotherapy [IMpower210] | Phase 3 | PD-L1 | Changchun (China), Daegu (Korea, Republic of), Daejeon (Korea, Republic of), Guangzhou (China), Hangzhou (China), Jeollanam-do (Korea, Republic of), Seoul (Korea, Republic of), Shanghai (China), Tianjin (天津) (China) | NCT02813785 |

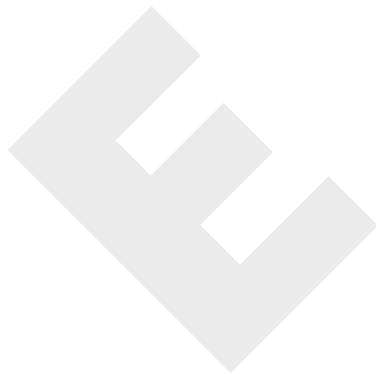

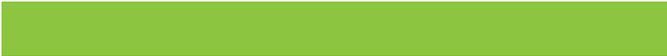

**NF2**
E427*

Inactivation or loss of NF2 results in the dysregulation of mTOR and FAK pathway signaling. Therefore, mTOR and/or FAK inhibitors may be relevant for patients with NF2 inactivating mutations.

Examples of clinical trials that may be appropriate for this patient are listed below. These trials were identified through a search of the trial website clinicaltrials.gov using keyword terms such as "NF2", "mTOR", "FAK", "everolimus", "temsirolimus", "GSK2256098", "VS-4718", "defactinib", "NSCLC", "lung", "solid tumor" and/or "advanced cancer".

| TITLE | PHASE | TARGETS | LOCATIONS | NCT ID |
|---|---|---|---|---|
| A Phase I Study of BKM120 and Everolimus in Advanced Solid Malignancies | Phase 1 | PI3K, mTOR | Georgia | NCT01470209 |
| A Phase I Study of VS-4718, a Focal Adhesion Kinase Inhibitor, in Subjects With Metastatic Non-Hematologic Malignancies | Phase 1 | FAK | Arizona, California, Florida, Tennessee | NCT01849744 |
| A Phase Ib/IIa Study of AZD2014 in Combination With Selumetinib in Patients With Advanced Cancers | Phase 1/Phase 2 | mTORC1, mTORC2, MEK | London (United Kingdom) | NCT02583542 |
| Phase II Study of Everolimus in Patients With Advanced Solid Malignancies With TSC1 and TSC2 Mutations | Phase 2 | mTOR | Missouri | NCT02352844 |
| A Multicenter, Open-label, Phase 1b Study of MLN0128 (an Oral mTORC1/2 Inhibitor) in Combination With MLN1117 (an Oral PI3Kα Inhibitor) in Adult Patients With Advanced Nonhematologic Malignancies | Phase 1 | PI3K-alpha, mTORC1, mTORC2 | Massachusetts, Tennessee, Texas, Barcelona (Spain), Sutton (United Kingdom) | NCT01899053 |

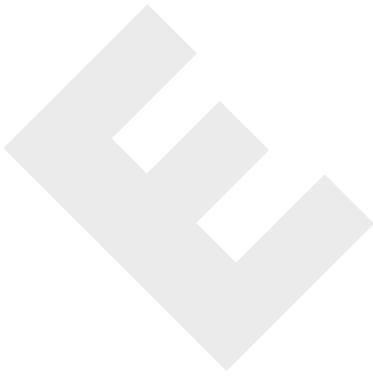

**For more comprehensive information please log on to the Interactive Cancer Explorer™**
**To set up your Interactive Cancer Explorer account, contact your sales representative or call 1-888-988-3639.**

STK11 loss or inactivating mutations may predict sensitivity to mTOR inhibitors or SRC inhibitors.

*STK11*

splice site 921-1G>C

Examples of clinical trials that may be appropriate for this patient are listed below. These trials were identified through a search of the trial website clinicaltrials.gov using keyword terms such as "mTOR", "SRC", "everolimus", "temsirolimus", "dasatinib", "bosutinib", "NSCLC", "lung", "solid tumor", and/or "advanced cancer".

| TITLE | PHASE | TARGETS | LOCATIONS | NCT ID |
|---|---|---|---|---|
| A Phase I Study of BKM120 and Everolimus in Advanced Solid Malignancies | Phase 1 | PI3K, mTOR | Georgia | NCT01470209 |
| A Multicenter, Open-label, Phase 1b Study of MLN0128 (an Oral mTORC1/2 Inhibitor) in Combination With MLN1117 (an Oral PI3Kα Inhibitor) in Adult Patients With Advanced Nonhematologic Malignancies | Phase 1 | PI3K-alpha, mTORC1, mTORC2 | Massachusetts, Tennessee, Texas, Barcelona (Spain), Sutton (United Kingdom) | NCT01899053 |
| Phase I Trial Evaluating Safety and Tolerability of the Irreversible Epidermal Growth Factor Receptor Inhibitor Afatinib (BIBW 2992) in Combination With the SRC Kinase Inhibitor Dasatinib for Patients With Non-small Cell Lung Cancer (NSCLC) | Phase 1 | EGFR, ERBB2, KIT, PDGFRs, SRC, ABL | Florida | NCT01999985 |
| Phase I Study of MLN0128 (NSC# 768435) in Combination With Ziv-Aflibercept (NSC# 724770) in Patients With Advanced Cancers | Phase 1 | mTORC1, mTORC2 | Texas | NCT02159989 |
| A Phase 1, Open-label Study to Evaluate the Safety, Tolerability, and Pharmacokinetics of MLN0128 (an Oral mTORC 1/2 Inhibitor) as a Single Agent and in Combination With Paclitaxel in Adult Patients With Advanced Nonhematologic Malignancies | Phase 1 | mTORC1, mTORC2 | Florida, Oklahoma, Tennessee | NCT02412722 |

**For more comprehensive information please log on to the Interactive Cancer Explorer™**
**To set up your Interactive Cancer Explorer account, contact your sales representative or call 1-888-988-3639.**

## APPENDIX

### VARIANTS OF UNKNOWN SIGNIFICANCE

Note: One or more variants of unknown significance (VUS) were detected in this patient's tumor. These variants may not have been adequately characterized in the scientific literature at the time this report was issued, and/or the genomic context of these alterations makes their significance unclear. We choose to include them here in the event that they become clinically meaningful in the future.

*AKT3*
E115K

*ARAF*
G446D

*BTK*
E215V

*CD79B*
splice site 592-2A>C

*CDK12*
amplification

*CDK8*
M273I

*CSF1R*
L471M

*EPHA5*
N120K

*ERBB4*
N1029K,splice site 3136-2A>T

*FLT4*
E1002K

*GLI1*
R1068G

*GNAS*
P98L

*GPR124*
Q227K,V688L

*HSD3B1*
V224_V226>G*CV, Y225C

*IL7R*
H309Q

*KEAP1*
E488K

*LRP1B*
D3844H,G4366W

*MDM2*
G137C

*Microsatellite status*
MS-Stable

*MLL2*
Q3738H

*MLL3*
C388F

*NOTCH1*
G661S

*NTRK3*
H632N

*PIK3CG*
L838M

*PRDM1*
amplification

*PRKDC*
E2175Q,rearrangement

*RUNX1T1*
D477Y,V16L

*SLIT2*
V1290A

*SMO*
V294F

*SPOP*
amplification

*WISP3*
amplification

**For more comprehensive information please log on to the Interactive Cancer Explorer™**
**To set up your Interactive Cancer Explorer account, contact your sales representative or call 1-888-988-3639.**

## APPENDIX

### GENES ASSAYED IN FOUNDATIONONE

FoundationOne is designed to include all genes known to be somatically altered in human solid tumors that are validated targets for therapy, either approved or in clinical trials, and/or that are unambiguous drivers of oncogenesis based on current knowledge. The current assay interrogates 315 genes as well as introns of 28 genes involved in rearrangements. The assay will be updated periodically to reflect new knowledge about cancer biology.

**DNA Gene List: Entire Coding Sequence for the Detection of Base Substitutions, Insertion/Deletions, and Copy Number Alterations**

| | | | | | | | | | |
|---|---|---|---|---|---|---|---|---|---|
| ABL1 | ABL2 | ACVR1B | AKT1 | AKT2 | AKT3 | ALK | AMER1 (FAM123B) | APC | AR |
| ARAF | ARFRP1 | ARID1A | ARID1B | ARID2 | ASXL1 | ATM | ATR | ATRX | AURKA |
| AURKB | AXIN1 | AXL | BAP1 | BARD1 | BCL2 | BCL2L1 | BCL2L2 | BCL6 | BCOR |
| BCORL1 | BLM | BRAF | BRCA1 | BRCA2 | BRD4 | BRIP1 | BTG1 | BTK | C11orf30 (EMSY) |
| CARD11 | CBFB | CBL | CCND1 | CCND2 | CCND3 | CCNE1 | CD274 | CD79A | CD79B |
| CDC73 | CDH1 | CDK12 | CDK4 | CDK6 | CDK8 | CDKN1A | CDKN1B | CDKN2A | CDKN2B |
| CDKN2C | CEBPA | CHD2 | CHD4 | CHEK1 | CHEK2 | CIC | CREBBP | CRKL | CRLF2 |
| CSF1R | CTCF | CTNNA1 | CTNNB1 | CUL3 | CYLD | DAXX | DDR2 | DICER1 | DNMT3A |
| DOT1L | EGFR | EP300 | EPHA3 | EPHA5 | EPHA7 | EPHB1 | ERBB2 | ERBB3 | ERBB4 |
| ERG | ERRFI1 | ESR1 | EZH2 | FAM46C | FANCA | FANCC | FANCD2 | FANCE | FANCF |
| FANCG | FANCL | FAS | FAT1 | FBXW7 | FGF10 | FGF14 | FGF19 | FGF23 | FGF3 |
| FGF4 | FGF6 | FGFR1 | FGFR2 | FGFR3 | FGFR4 | FH | FLCN | FLT1 | FLT3 |
| FLT4 | FOXL2 | FOXP1 | FRS2 | FUBP1 | GABRA6 | GATA1 | GATA2 | GATA3 | GATA4 |
| GATA6 | GID4 (C17orf39) | GLI1 | GNA11 | GNA13 | GNAQ | GNAS | GPR124 | GRIN2A | GRM3 |
| GSK3B | H3F3A | HGF | HNF1A | HRAS | HSD3B1 | HSP90AA1 | IDH1 | IDH2 | IGF1R |
| IGF2 | IKBKE | IKZF1 | IL7R | INHBA | INPP4B | IRF2 | IRF4 | IRS2 | JAK1 |
| JAK2 | JAK3 | JUN | KAT6A (MYST3) | KDM5A | KDM5C | KDM6A | KDR | KEAP1 | KEL |
| KIT | KLHL6 | KMT2A (MLL) | KMT2C (MLL3) | KMT2D (MLL2) | KRAS | LMO1 | LRP1B | LYN | LZTR1 |
| MAGI2 | MAP2K1 | MAP2K2 | MAP2K4 | MAP3K1 | MCL1 | MDM2 | MDM4 | MED12 | MEF2B |
| MEN1 | MET | MITF | MLH1 | MPL | MRE11A | MSH2 | MSH6 | MTOR | MUTYH |
| MYC | MYCL (MYCL1) | MYCN | MYD88 | NF1 | NF2 | NFE2L2 | NFKBIA | NKX2-1 | NOTCH1 |
| NOTCH2 | NOTCH3 | NPM1 | NRAS | NSD1 | NTRK1 | NTRK2 | NTRK3 | NUP93 | PAK3 |
| PALB2 | PARK2 | PAX5 | PBRM1 | PDCD1LG2 | PDGFRA | PDGFRB | PDK1 | PIK3C2B | PIK3CA |
| PIK3CB | PIK3CG | PIK3R1 | PIK3R2 | PLCG2 | PMS2 | POLD1 | POLE | PPP2R1A | PRDM1 |
| PREX2 | PRKAR1A | PRKCI | PRKDC | PRSS8 | PTCH1 | PTEN | PTPN11 | QKI | RAC1 |
| RAD50 | RAD51 | RAF1 | RANBP2 | RARA | RB1 | RBM10 | RET | RICTOR | RNF43 |
| ROS1 | RPTOR | RUNX1 | RUNX1T1 | SDHA | SDHB | SDHC | SDHD | SETD2 | SF3B1 |
| SLIT2 | SMAD2 | SMAD3 | SMAD4 | SMARCA4 | SMARCB1 | SMO | SNCAIP | SOCS1 | SOX10 |
| SOX2 | SOX9 | SPEN | SPOP | SPTA1 | SRC | STAG2 | STAT3 | STAT4 | STK11 |
| SUFU | SYK | TAF1 | TBX3 | TERC | TERT (promoter only) | TET2 | TGFBR2 | TNFAIP3 | TNFRSF14 |
| TOP1 | TOP2A | TP53 | TSC1 | TSC2 | TSHR | U2AF1 | VEGFA | VHL | WISP3 |
| WT1 | XPO1 | ZBTB2 | ZNF217 | ZNF703 | | | | | |

**DNA Gene List: For the Detection of Select Rearrangements**

| | | | | | | | | | |
|---|---|---|---|---|---|---|---|---|---|
| ALK | BCL2 | BCR | BRAF | BRCA1 | BRCA2 | BRD4 | EGFR | ETV1 | ETV4 |
| ETV5 | ETV6 | FGFR1 | FGFR2 | FGFR3 | KIT | MSH2 | MYB | MYC | NOTCH2 |
| NTRK1 | NTRK2 | PDGFRA | RAF1 | RARA | RET | ROS1 | TMPRSS2 | | |

**Additional Assays: For the Detection of Select Cancer Biomarkers**

*Microsatellite status*

*Tumor Mutation Burden*

**For more comprehensive information please log on to the Interactive Cancer Explorer™**
**To set up your Interactive Cancer Explorer account, contact your sales representative or call 1-888-988-3639.**

## FOUNDATIONONE PERFORMANCE SPECIFICATIONS

| ACCURACY | | |
|---|---|---|
| **Sensitivity: Base Substitutions** | At Mutant Allele Frequency ≥10% | >99.9%  (CI* 99.6%-100%) |
| | At Mutant Allele Frequency 5-10% | 99.3%  (CI* 98.3%-99.8%) |
| **Sensitivity: Insertions/Deletions (1-40 bp)** | At Mutant Allele Frequency ≥20% | 97.9%  (CI* 92.5%-99.7%) |
| | At Mutant Allele Frequency 10-20% | 97.3%  (CI* 90.5%-99.7%) |
| **Sensitivity: Copy Number Alterations—Amplifications** (ploidy <4, Amplification with  Copy Number ≥8) | At ≥30% tumor nuclei | >99.0%  (CI* 93.6%-100%) |
| | At  20% tumor nuclei | 92.6%  (CI* 66.1%-99.8%) |
| **Sensitivity: Copy Number Alterations—Deletions** (ploidy <4, Homozygous Deletions) | At ≥30% tumor nuclei | 97.2%  (CI* 85.5%-99.9%) |
| | At  20% tumor nuclei | 88.9%  (CI* 51.8%-99.7%) |
| **Sensitivity: Rearrangements** (selected rearrangements in specimens with ≥20% tumor nuclei)** | | >90.0% [1] <br> >99.0%  for ALK fusion[2] <br> (CI* 89.1%-100%) |
| **Sensitivity: Microsatellite status** | At ≥20% tumor nuclei | 97.0%  (CI* 89.6%-99.6%) |
| **Specificity: all variant types** | Positive Predictive Value (PPV) | >99.0% |
| **Specificity: Microsatellite status** | Positive Predictive Value (PPV) | >95.0% |
| **Accuracy: Tumor Mutation Burden** | At ≥20% tumor nuclei | >90.0% |
| **REPRODUCIBILITY** (average concordance between replicates) | | 96.4% inter-batch precision <br> 98.9% intra-batch precision <br> 95.8% microsatellite status precision <br> 96.4% tumor mutation burden precision |

* 95% Confidence Interval

** Performance for gene fusions within targeted introns only. Sensitivity for gene fusions occurring outside targeted introns or in highly repetitive intronic sequence contexts is reduced.

[1] Based on analysis of coverage and rearrangement structure in the COSMIC database for the solid tumor fusion genes where alteration prevalence could be established, complemented by detection of exemplar rearrangements in cell line titration experiments.

[2] Based on ALK rearrangement concordance analysis vs. a standard clinical FISH assay described in: Yelensky, R. et al. Analytical validation of solid tumor fusion gene detection in a comprehensive NGS-based clinical cancer genomic test, In: Proceedings of the 105th Annual Meeting of the American Association for Cancer Research; 2014 Apr 5-9; San Diego, CA. Philadelphia (PA): AACR; 2014. Abstract nr 4699

Assay specifications were determined for typical median exon coverage of approximately 500X. For additional information regarding the validation of FoundationOne, please refer to the article, Frampton, GM. et al. Development and validation of a clinical cancer genomic profiling test based on massively parallel DNA sequencing, Nat Biotechnol (2013 Oct. 20).

Microsatellite status (a measure of microsatellite instability, or "MSI") is determined by assessing indel characteristics at 114 homopolymer repeat loci in or near the targeted gene regions of the FoundationOne test. Microsatellite status is assayed for all FoundationOne samples. MSI-High results are reported in all tumor types. In select tumor types, other Microsatellite status results may be reported (MS-Stable, MSI-Ambiguous, MSI-Unknown) when relevant. Microsatellite status result may be reported as "Unknown" if the sample is not of sufficient quality to confidently determine Microsatellite status.

Tumor Mutation Burden (TMB) is determined by measuring the number of somatic mutations occurring in sequenced genes on the FoundationOne and FoundationOne Heme tests and extrapolating to the genome as a whole. TMB is assayed for all FoundationOne and FoundationOne Heme samples. TMB-High results are reported in all tumor types.  In select tumor types, other TMB results may be reported (TMB-Intermediate, TMB-Low, TMB-Unknown) when relevant. TMB results are determined as follows: TMB-High corresponds to greater than or equal to 20 mutations per megabase (Muts/Mb); TMB-Intermediate corresponds to 6-19 Muts/Mb; TMB-Low corresponds to less than or equal to 5 Muts/Mb. Tumor Mutation Burden may be reported as "Unknown" if the sample is not of sufficient quality to confidently determine Tumor Mutation Burden.

For additional information specific to the performance of this specimen, please contact Foundation Medicine, Inc. at 1-888-988-3639.

**For more comprehensive information please log on to the Interactive Cancer Explorer™**
**To set up your Interactive Cancer Explorer account, contact your sales representative or call  1-888-988-3639.**

**For more comprehensive information please log on to the Interactive Cancer Explorer™**
**To set up your Interactive Cancer Explorer account, contact your sales representative or call  1-888-988-3639.**

**For more comprehensive information please log on to the Interactive Cancer Explorer™**
**To set up your Interactive Cancer Explorer account, contact your sales representative or call 1-888-988-3639.**

**For more comprehensive information please log on to the Interactive Cancer Explorer™**
**To set up your Interactive Cancer Explorer account, contact your sales representative or call  1-888-988-3639.**

**For more comprehensive information please log on to the Interactive Cancer Explorer™**
**To set up your Interactive Cancer Explorer account, contact your sales representative or call  1-888-988-3639.**

**For more comprehensive information please log on to the Interactive Cancer Explorer™**
**To set up your Interactive Cancer Explorer account, contact your sales representative or call 1-888-988-3639.**

**For more comprehensive information please log on to the Interactive Cancer Explorer™**
**To set up your Interactive Cancer Explorer account, contact your sales representative or call 1-888-988-3639.**

**For more comprehensive information please log on to the Interactive Cancer Explorer™**
**To set up your Interactive Cancer Explorer account, contact your sales representative or call  1-888-988-3639.**

**For more comprehensive information please log on to the Interactive Cancer Explorer™**
**To set up your Interactive Cancer Explorer account, contact your sales representative or call 1-888-988-3639.**

**For more comprehensive information please log on to the Interactive Cancer Explorer™**
**To set up your Interactive Cancer Explorer account, contact your sales representative or call 1-888-988-3639.**

**For more comprehensive information please log on to the Interactive Cancer Explorer™**
**To set up your Interactive Cancer Explorer account, contact your sales representative or call 1-888-988-3639.**

**For more comprehensive information please log on to the Interactive Cancer Explorer™**
**To set up your Interactive Cancer Explorer account, contact your sales representative or call 1-888-988-3639.**

**For more comprehensive information please log on to the Interactive Cancer Explorer™**
**To set up your Interactive Cancer Explorer account, contact your sales representative or call 1-888-988-3639.**

**For more comprehensive information please log on to the Interactive Cancer Explorer™**
**To set up your Interactive Cancer Explorer account, contact your sales representative or call  1-888-988-3639.**

## ABOUT FOUNDATIONONE™

**FoundationOne™:** FoundationOne was developed and its performance characteristics determined by Foundation Medicine, Inc. (Foundation Medicine). FoundationOne has not been cleared or approved by the United States Food and Drug Administration (FDA). The FDA has determined that such clearance or approval is not necessary. FoundationOne may be used for clinical purposes and should not be regarded as purely investigational or for research only. Foundation Medicine's clinical reference laboratory is certified under the Clinical Laboratory Improvement Amendments of 1988 (CLIA) as qualified to perform high-complexity clinical testing.

**Diagnostic Significance:** FoundationOne identifies alterations to select cancer-associated genes or portions of genes (biomarkers). In some cases, the Test Report also highlights selected negative test results regarding biomarkers of clinical significance.

**Qualified Alteration Calls (Equivocal and Subclonal):** An alteration denoted as "amplification – equivocal" implies that the FoundationOne assay data provide some, but not unambiguous, evidence that the copy number of a gene exceeds the threshold for identifying copy number amplification. The threshold used in FoundationOne for identifying a copy number amplification is five (5) for ERBB2 and six (6) for all other genes. Conversely, an alteration denoted as "loss – equivocal" implies that the FoundationOne assay data provide some, but not unambiguous, evidence for homozygous deletion of the gene in question. An alteration denoted as "subclonal" is one that the FoundationOne analytical methodology has identified as being present in <10% of the assayed tumor DNA.

**The Report** incorporates analyses of peer-reviewed studies and other publicly available information identified by Foundation Medicine; these analyses and information may include associations between a molecular alteration (or lack of alteration) and one or more drugs with potential clinical benefit (or potential lack of clinical benefit), including drug candidates that are being studied in clinical research.

**NOTE:** A finding of biomarker alteration does not necessarily indicate pharmacologic effectiveness (or lack thereof) of any drug or treatment regimen; a finding of no biomarker alteration does not necessarily indicate lack of pharmacologic effectiveness (or effectiveness) of any drug or treatment regimen.

**Alterations and Drugs Not Presented in Ranked Order:** In this Report, neither any biomarker alteration, nor any drug associated with potential clinical benefit (or potential lack of clinical benefit), are ranked in order of potential or predicted efficacy.

**Level of Evidence Not Provided:** Drugs with potential clinical benefit (or potential lack of clinical benefit) are not evaluated for source or level of published evidence.

**No Guarantee of Clinical Benefit:** This Report makes no promises or guarantees that a particular drug will be effective in the treatment of disease in any patient. This Report also makes no promises or guarantees that a drug with potential lack of clinical benefit will in fact provide no clinical benefit.

**No Guarantee of Reimbursement:** Foundation Medicine makes no promises or guarantees that a healthcare provider, insurer or other third party payor, whether private or governmental, will reimburse a patient for the cost of FoundationOne.

**Treatment Decisions are Responsibility of Physician:** Drugs referenced in this Report may not be suitable for a particular patient. The selection of any, all or none of the drugs associated with potential clinical benefit (or potential lack of clinical benefit) resides entirely within the discretion of the treating physician. Indeed, the information in this Report must be considered in conjunction with all other relevant information regarding a particular patient, before the patient's treating physician recommends a course of treatment.

Decisions on patient care and treatment must be based on the independent medical judgment of the treating physician, taking into consideration all applicable information concerning the patient's condition, such as patient and family history, physical examinations, information from other diagnostic tests, and patient preferences, in accordance with the standard of care in a given community. A treating physician's decisions should not be based on a single test, such as this Test, or the information contained in this Report.

Certain sample or variant characteristics may result in reduced sensitivity. These include: subclonal alterations in heterogeneous samples, low sample quality or with homozygous losses of <3 exons; and deletions and insertions >40bp, or in repetitive/high homology sequences. FoundationOne is performed using DNA derived from tumor, and as such germline events may not be reported. The following targets typically have low coverage resulting in a reduction in sensitivity: *SDHD* exon 6 and *TP53* exon 1.

FoundationOne complies with all European Union (EU) requirements of the IVD Directive 98/79EC. As such, the FoundationOne Assay has been registered for CE mark by our EU Authorized Representative, Qarad b.v.b.a, Cipalstraat 3, 2440 Geel, Belgium.

**For more comprehensive information please log on to the Interactive Cancer Explorer™**
**To set up your Interactive Cancer Explorer account, contact your sales representative or call 1-888-988-3639.**

# SUPPLEMENTARY FIGURE S2

# MI PROFILE™
MOLECULAR INTELLIGENCE TUMOR REPORT

CARIS LIFE SCIENCES

## FINAL REPORT

### PATIENT
**Name:** Patient, Test
**Date of Birth:** XX-Mon-1932
**Sex:** Male
**Case Number:** TN14-111111
**Diagnosis:** Adenocarcinoma, NOS

### SPECIMEN INFORMATION
**Primary Tumor Site:** Lung, NOS
**Specimen Site:** Lung, NOS
**Specimen ID:** ABC-12345-YZ
**Specimen Collected:** XX-Mon-2014
**Completion of Testing:** XX-Mon-2014

### ORDERED BY
**Ordering Physician, MD**
**The Cancer Center**
123 Main Street
Springfield, XY 12345
(123) 456-7890

**Bold Therapies** = On NCCN Compendium® Therapies

---

## ✔ THERAPIES WITH **POTENTIAL BENEFIT** (PAGE 4)

| | | | | | |
|---|---|---|---|---|---|
| **afatinib** | EGFR★ | irinotecan | TOPO1 | dacarbazine, temozolomide | MGMT★ |
| **docetaxel, nab-paclitaxel, paclitaxel** | TLE3★, PGP | **pemetrexed** | TS★ | doxorubicin, epirubicin, liposomal-doxorubicin | PGP, TOP2A |
| **erlotinib** | PTEN, PIK3CA, KRAS, EGFR★ | capecitabine, fluorouracil | TS★ | gefitinib | PTEN, PIK3CA, KRAS, EGFR★ |
| **gemcitabine** | RRM1★ | | | | |

★ Indicates Clinical Trial Opportunity • 282 Chemotherapy Trials • 102 Targeted Therapy Trials  *(See Clinical Trials Connector™ on page 8 for details.)*

---

## ✘ THERAPIES WITH **POTENTIAL LACK OF BENEFIT** (PAGE 6)

| | | | | | |
|---|---|---|---|---|---|
| **ceritinib** | ALK | **dabrafenib, vemurafenib** | BRAF | cetuximab | EGFR |
| **crizotinib** | ROS1, ALK | | | | |

---

## ? THERAPIES WITH **INDETERMINATE BENEFIT** (PAGE 7)

| | | |
|---|---|---|
| **trastuzumab†** | everolimus, temsirolimus | lapatinib† |
| ado-trastuzumab emtansine (T-DM1)†, pertuzumab† | imatinib | vandetanib |

†Association to Benefit was not indicated due to assay failure.

---

Therapies associated with potential benefit or lack of benefit, as indicated above, are based on biomarker results provided in this report and are based on published medical evidence. This evidence may have been obtained from studies performed in the cancer type present in the tested patient's sample or derived from another tumor type. The selection of any, all, or none of the matched therapies resides solely with the discretion of the treating physician. Decisions on patient care and treatment must be based on the independent medical judgment of the treating physician, taking into consideration all available information in addition to this report concerning the patient's condition in accordance with the applicable standard of care.

CE

| Biomarker | Method | Result | Biomarker | Method | Result |
|-----------|--------|--------|-----------|--------|--------|
| ABL1 | NGS | Mutation Not Detected | KDR (VEGFR2) | NGS | Mutation Not Detected |
| AKT1 | NGS | Quantity Not Sufficient | KRAS | NGS | Mutation Not Detected |
| ALK | FISH | Negative | MGMT | IHC | Negative |
| ALK | NGS | Mutation Not Detected | MPL | NGS | Mutation Not Detected |
| Androgen Receptor | IHC | Negative | NOTCH1 | NGS | Mutation Not Detected |
| APC | NGS | Mutation Not Detected | NPM1 | NGS | Mutation Not Detected |
| ATM | NGS | Mutation Not Detected | NRAS | NGS | Mutation Not Detected |
| BRAF | NGS | Mutation Not Detected | PD-1 | IHC | Negative |
| CDH1 | NGS | Mutation Not Detected | PDGFRA | NGS | Mutation Not Detected |
| c-KIT | NGS | Mutation Not Detected | PD-L1 | IHC | Negative |
| cMET | CISH | Test Not Performed | PGP | IHC | Negative |
| cMET | NGS | Mutation Not Detected | PIK3CA | NGS | Mutation Not Detected |
| cMET | IHC | Negative | PR | IHC | Negative |
| CSF1R | NGS | Mutation Not Detected | PTEN | NGS | Mutation Not Detected |
| CTNNB1 | NGS | Mutation Not Detected | PTEN | IHC | Positive |
| EGFR | NGS | Mutated, Pathogenic Exon 21 \| L858R | PTPN11 | NGS | Mutation Not Detected |
| EGFR | IHC (H-Score) | Negative | RB1 | NGS | Mutation Not Detected |
| ER | IHC | Negative | RET | NGS | Mutation Not Detected |
| ERBB4 | NGS | Mutation Not Detected | ROS1 | FISH | Negative |
| FBXW7 | NGS | Mutation Not Detected | RRM1 | IHC | Negative |
| FGFR1 | NGS | Mutation Not Detected | SMAD4 | NGS | Mutation Not Detected |
| FGFR2 | NGS | Mutation Not Detected | SMARCB1 | NGS | Mutation Not Detected |
| FLT3 | NGS | Mutation Not Detected | SMO | NGS | Quantity Not Sufficient |
| GNA11 | NGS | Quantity Not Sufficient | SPARC Monoclonal | IHC | Negative |
| GNAQ | NGS | Mutation Not Detected | SPARC Polyclonal | IHC | Positive |
| GNAS | NGS | Mutation Not Detected | STK11 | NGS | Quantity Not Sufficient |
| Her2/Neu | CISH | Test Not Performed | TLE3 | IHC | Positive |
| Her2/Neu | IHC | Negative | TOP2A | IHC | Positive |
| Her2/Neu (ERBB2) | NGS | Mutation Not Detected | TOPO1 | IHC | Positive |
| HNF1A | NGS | Mutation Not Detected | TP53 | NGS | Mutated, Presumed Pathogenic Exon 5 \| V173L |
| HRAS | NGS | Quantity Not Sufficient | TS | IHC | Negative |
| IDH1 | NGS | Mutation Not Detected | TUBB3 | IHC | Positive |
| JAK2 | NGS | Mutation Not Detected | VHL | NGS | Quantity Not Sufficient |
| JAK3 | NGS | Mutation Not Detected | | | |

**FISH:** Fluorescence *in situ* hybridization   **IHC:** Immunohistochemistry   **CISH:** Chromogenic *in situ* hybridization   **NGS:** Next-Generation Sequencing

| PATIENT: Patient, Test (XX-Mon-1932) | TN14-111111 | PHYSICIAN: Ordering Physician, MD |
|---|---|---|

## SUMMARY OF BIOMARKER RESULTS (SEE APPENDIX FOR FULL DETAILS)

The Next-Generation Sequencing results above include only the genes most commonly associated with cancer. See summary below and for full Next-Generation Sequencing results, see Appendix page 1.
Genes tested: 44 | Genes with actionable mutations: 2 | Genes with unclassified mutations: 0 | Genes with no mutations detected: 36

See the Appendix section for a detailed overview of the biomarker test results for each technology.

✔ THERAPIES WITH **POTENTIAL BENEFIT**

| Therapies | Test | Method | Result | Value† | Clinical Association | | | | | |
|-----------|------|--------|--------|--------|---------------------|--|--|--|--|--|
| | | | | | Potential Benefit | Decreased Potential Benefit | Lack of Potential Benefit | Highest Level of Evidence* | Reference |
| afatinib | EGFR | Next Gen SEQ | Mutated, Pathogenic | L858R | ✔ | | | I / Good | 4# |
| capecitabine, fluorouracil, pemetrexed | TS | IHC | Negative | 1+ 1% | ✔ | | | II-1 / Good | 5#, 6#, 7 |
| dacarbazine, temozolomide | MGMT | IHC | Negative | 1+ 10% | ✔ | | | II-2 / Good | 19, 20 |
| docetaxel, nab-paclitaxel, paclitaxel | PGP | IHC | Negative | 0+ 100% | ✔ | | | II-3 / Fair | 22, 23# |
| | TLE3 | IHC | Positive | 2+ 30% | ✔ | | | II-2 / Good | 21 |
| | TUBB3 | IHC | Positive | 3+ 90% | | ✔ | | I / Good | 24, 25#, 26#, 27# |
| doxorubicin, epirubicin, liposomal-doxorubicin | Her2/Neu | CISH | Technical Issues | | | | | | |
| | PGP | IHC | Negative | 0+ 100% | ✔ | | | II-1 / Fair | 28, 29 |
| | TOP2A | IHC | Positive | 2+ 10% | ✔ | | | I / Good | 30, 31 |
| erlotinib, gefitinib | cMET | CISH | Technical Issues | | | | | | |
| | EGFR | Next Gen SEQ | Mutated, Pathogenic | L858R | ✔ | | | I / Good | 35#, 37#, 38#, 39# |
| | KRAS | Next Gen SEQ | Mutation Not Detected | | ✔ | | | I / Good | 35#, 36# |
| | PIK3CA | Next Gen SEQ | Mutation Not Detected | | ✔ | | | II-1 / Good | 33#, 34# |
| | PTEN | IHC | Positive | 2+ 95% | ✔ | | | II-3 / Fair | 32# |
| gemcitabine | RRM1 | IHC | Negative | 2+ 15% | ✔ | | | I / Good | 43# |

*Additional Therapies Associated with Potential Benefit continued on the next page. >*

## ✔ THERAPIES WITH **POTENTIAL BENEFIT**

| Therapies | Test | Method | Result | Value† | Clinical Association | | | | |
|-----------|------|--------|--------|--------|---------------------|---|---|---|---|
| | | | | | Potential Benefit | Decreased Potential Benefit | Lack of Potential Benefit | Highest Level of Evidence* | Reference |
| irinotecan | TOPO1 | IHC | Positive | 2+ 80% | ✔ | | | II-1 / Good | 49, 50, 51 |

\* The level of evidence for all references is assigned according to the Literature Level of Evidence Framework consistent with the US Preventive Services Task Force described further in the Appendix of this report. The data level of each biomarker-drug interaction is the highest level of evidence based on the body of evidence, overall clinical utility, competing biomarker interactions and tumor type from which the evidence was gathered.

\# Evidence reference includes data from the same lineage as the tested specimen.

**†Refer to Appendix for detailed Result and Value information for each biomarker, including appropriate cutoffs, unit of measure, etc.**

## ✖ THERAPIES WITH **POTENTIAL LACK OF BENEFIT**

| Therapies | Test | Method | Result | Value† | Clinical Association | | | | |
|-----------|------|--------|--------|--------|---------------------|---|---|---|---|
| | | | | | Potential Benefit | Decreased Potential Benefit | Lack of Potential Benefit | Highest Level of Evidence* | Reference |
| ceritinib | ALK | FISH | Negative | | | | ✔ | II-1 / Good | 8# |
| cetuximab | EGFR | IHC H-Score | Negative | 180 | | | ✔ | I / Good | 9# |
| crizotinib | ALK | FISH | Negative | | | | ✔ | I / Good | 13#, 14# |
| | ROS1 | FISH | Negative | | | | ✔ | III / Good | 10#, 11#, 12# |
| dabrafenib, vemurafenib | BRAF | Next Gen SEQ | Mutation Not Detected | | | | ✔ | I / Good | 15#, 16, 17#, 18 |

\* The level of evidence for all references is assigned according to the Literature Level of Evidence Framework consistent with the US Preventive Services Task Force described further in the Appendix of this report. The data level of each biomarker-drug interaction is the highest level of evidence based on the body of evidence, overall clinical utility, competing biomarker interactions and tumor type from which the evidence was gathered.

\# Evidence reference includes data from the same lineage as the tested specimen.

**†Refer to Appendix for detailed Result and Value information for each biomarker, including appropriate cutoffs, unit of measure, etc.**

## ? THERAPIES WITH INDETERMINATE BENEFIT
(Biomarker results do not impact potential benefit or lack of potential benefit)

| Therapies | Test | Method | Result | Value† | Clinical Association | | | | |
|-----------|------|--------|--------|--------|---------------------|--|--|--|--|
| | | | | | Potential Benefit | Decreased Potential Benefit | Lack of Potential Benefit | Highest Level of Evidence* | Reference |
| ado-trastuzumab emtansine (T-DM1), pertuzumab | Her2/Neu | CISH | Technical Issues | | | | | | |
| | Her2/Neu | IHC | Negative | 0+ 100% | | | ✔ | I / Good | 1, 2, 3 |
| everolimus, temsirolimus | PIK3CA | Next Gen SEQ | Mutation Not Detected | | | ✔ | | II-2 / Good | 40, 41#, 42 |
| imatinib | c-KIT | Next Gen SEQ | Mutation Not Detected | | | | ✔ | II-2 / Good | 44, 45 |
| | PDGFRA | Next Gen SEQ | Mutation Not Detected | | | | ✔ | II-3 / Good | 46, 47, 48 |
| lapatinib | Her2/Neu | CISH | Technical Issues | | | | | | |
| | Her2/Neu | IHC | Negative | 0+ 100% | | | ✔ | I / Good | 52, 53, 54 |
| trastuzumab | Her2/Neu | CISH | Technical Issues | | | | | | |
| | Her2/Neu | IHC | Negative | 0+ 100% | | | ✔ | I / Good | 57, 58, 59, 60 |
| | Her2/Neu (ERBB2) | Next Gen SEQ | Mutation Not Detected | | | | ✔ | II-3 / Good | 55#, 56# |
| vandetanib | RET | Next Gen SEQ | Mutation Not Detected | | | | | I / Good | 61 |

\* The level of evidence for all references is assigned according to the Literature Level of Evidence Framework consistent with the US Preventive Services Task Force described further in the Appendix of this report. The data level of each biomarker-drug interaction is the highest level of evidence based on the body of evidence, overall clinical utility, competing biomarker interactions and tumor type from which the evidence was gathered.

\# Evidence reference includes data from the same lineage as the tested specimen.

†Refer to Appendix for detailed Result and Value information for each biomarker, including appropriate cutoffs, unit of measure, etc.

## CLINICAL TRIALS CONNECTOR™

For a complete list of open, enrolling clinical trials visit MI Portal to access the **Clinical Trials Connector** . This personalized, real-time web-based service provides additional clinical trial information and enhanced searching capabilities, including, but not limited to:

- Location: filter by geographic area
- Biomarker(s): identify specific biomarkers associated with open clinical trials to choose from
- Drug(s): search for specific therapies
- Trial Sponsor: locate trials based on the organization supporting the trial(s)

**Visit www.CarisMolecularIntelligence.com to view all matched trials.**

| CHEMOTHERAPY CLINICAL TRIALS (282) | | | |
|---|---|---|---|
| **Drug Class** | **Biomarker** | **Method** | **Investigational Agent(s)** |
| Alkylating agents (7) | MGMT | IHC | dacarbazine, temozolomide |
| Antifolates (64) | TS | IHC | methotrexate, pemetrexed |
| Nanoparticle-bound agents (23) | SPARC Polyclonal | IHC | nab-paclitaxel |
| Nucleoside analog (53) | RRM1 | IHC | gemcitabine |
| Pyrimidine analog (17) | TS | IHC | capecitabine, fluorouracil |
| Taxanes (118) | TLE3 | IHC | docetaxel, paclitaxel |

| TARGETED THERAPY CLINICAL TRIALS (102) | | | |
|---|---|---|---|
| **Drug Class** | **Biomarker** | **Method** | **Investigational Agent(s)** |
| Cell cycle inhibitors (5) | TP53 | Next Gen SEQ | MK-1775 |
| EGFR TKIs (77) | EGFR | Next Gen SEQ | CO-1686, erlotinib, gefitinib |
| Pan-HER inhibitors (20) | EGFR | Next Gen SEQ | afatinib, dacomitinib, icotinib, neratinib |

( ) = represents the total number of clinical trials identified by the Clinical Trials Connector for the provided drug class or table.

| **PATIENT:** Patient, Test (XX-Mon-1932) | **TN14-111111** | **PHYSICIAN:** Ordering Physician, MD |
|---|---|---|

# MI PROFILE™
MOLECULAR INTELLIGENCE TUMOR REPORT

# CARIS
LIFE SCIENCES®

**To view the rest of the report, contact a Caris Molecular Intelligence™ representative today.**

**(888) 979-8669**
**MIclientservices@carisls.com**

# SUPPLEMENTARY FIGURE S3

# Precision Oncology Reports Survey Results

**Figure S3.1.**

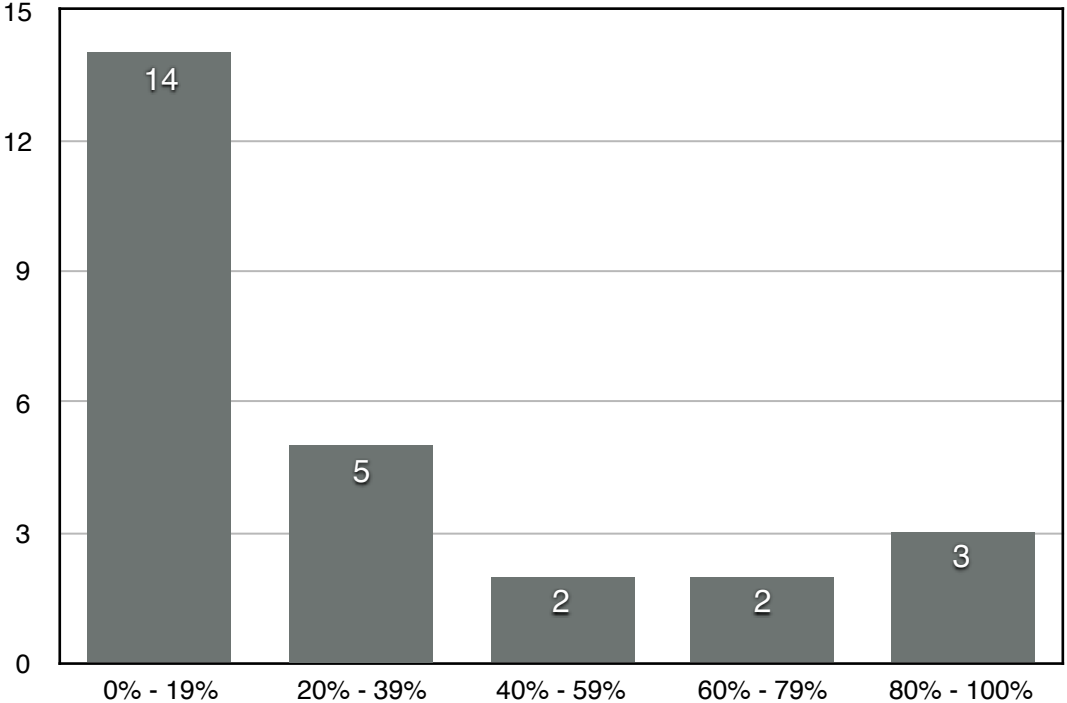

In the past 12 months, for what percentage of your patients did you order tumor genomic testing? (n = 26)

**Figure S3.2.**

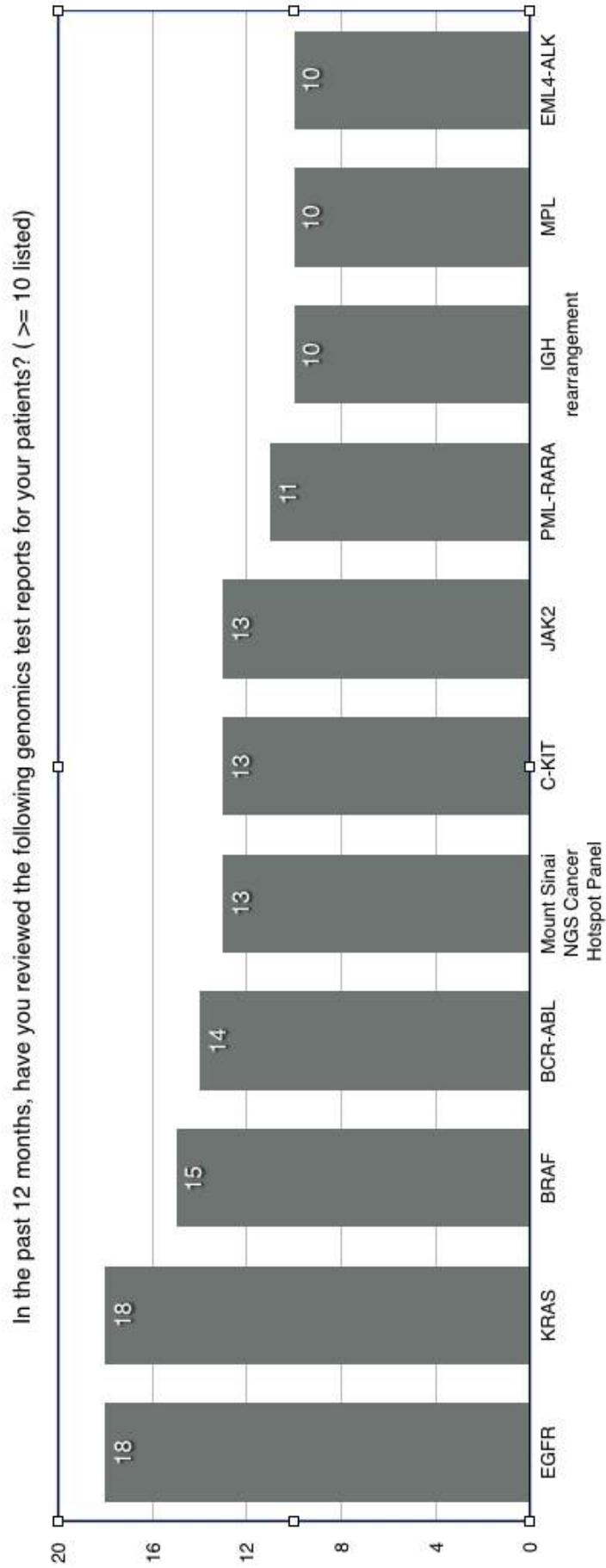

In the past 12 months, have you reviewed the following genomics test reports for your patients? ( >= 10 listed)

**Figure S3.3.**

Which of the following vendor genomic test reports have you used?
(n = 29, multiple answers allowed)

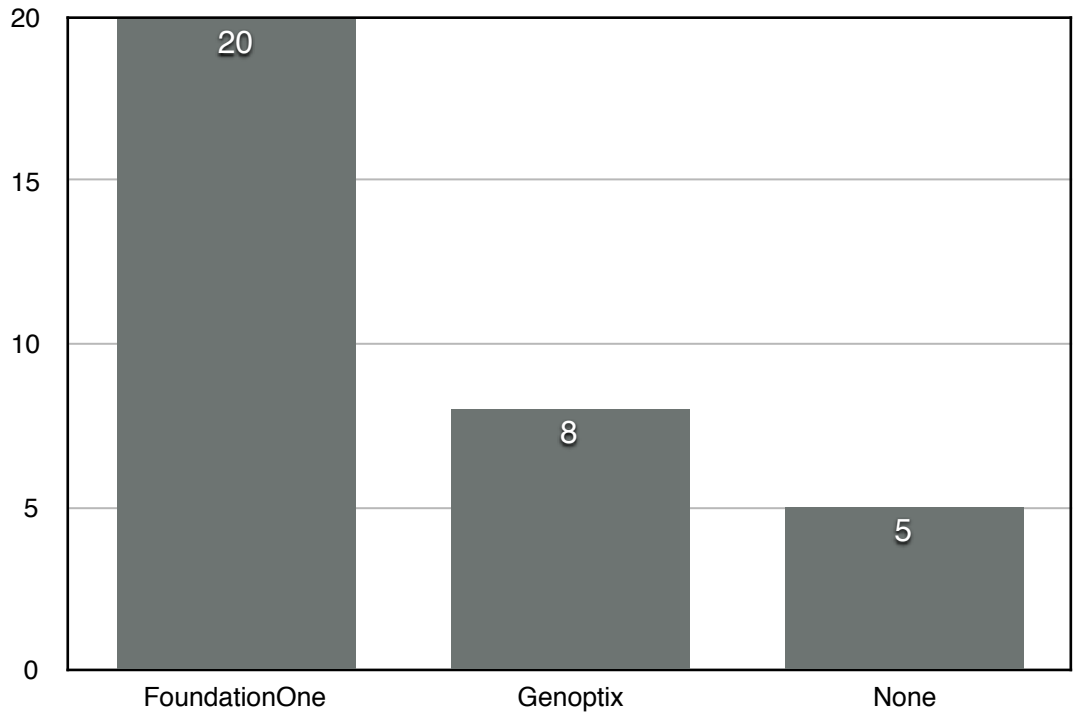

**Figure S3.4.**

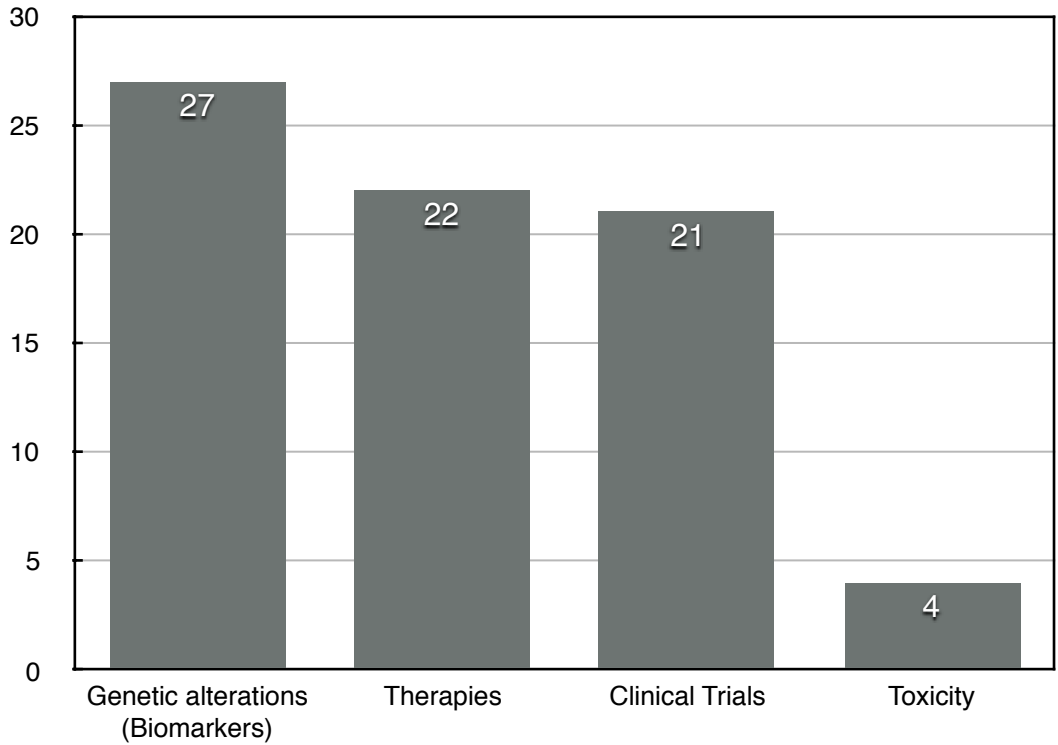

What is the most important information you expect to find in a tumor profiling report? (n = 32, multiple answers allowed)

**Figure S3.5.**

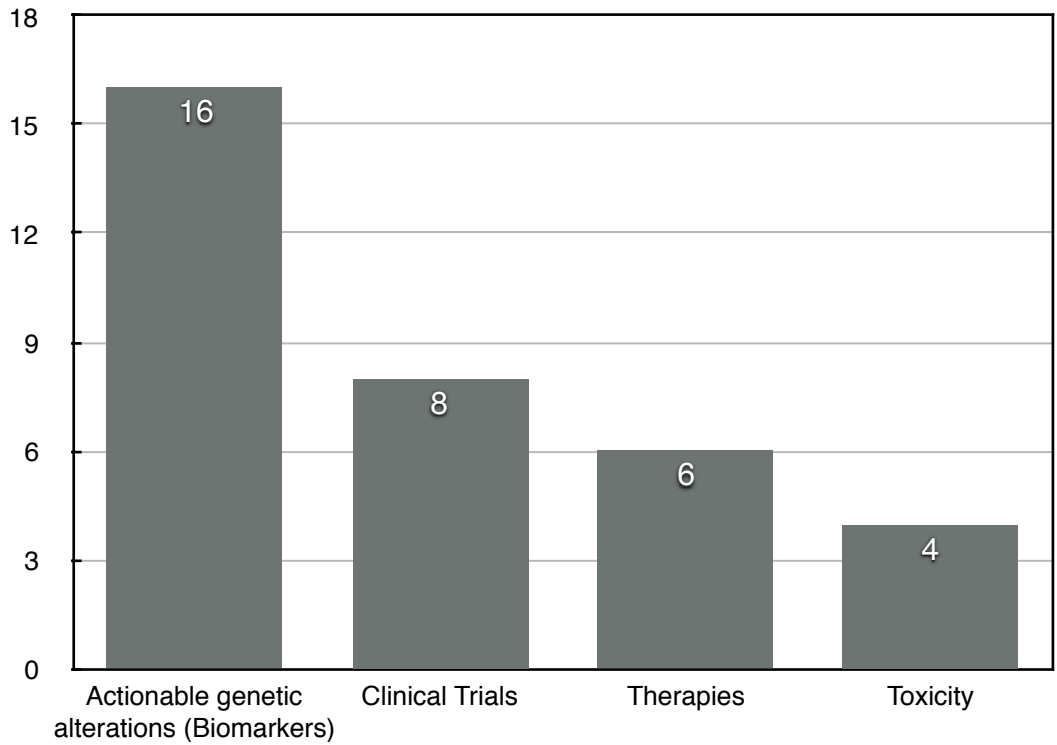

What is the most time-consuming component of interpreting a tumor profiling report in your experience? (n = 28, multiple answers allowed)

**Figure S3.6.**

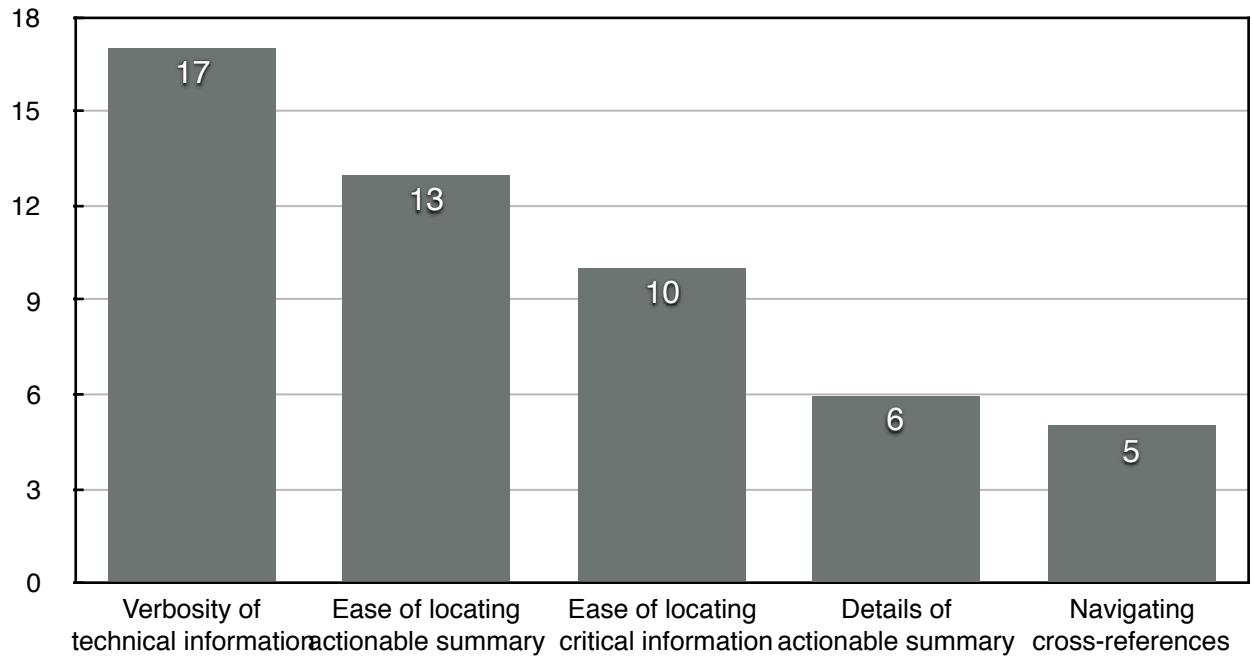

What parts are most troublesome reading in these reports? (n = 29, multiple answers allowed)

**Figure S3.7.**

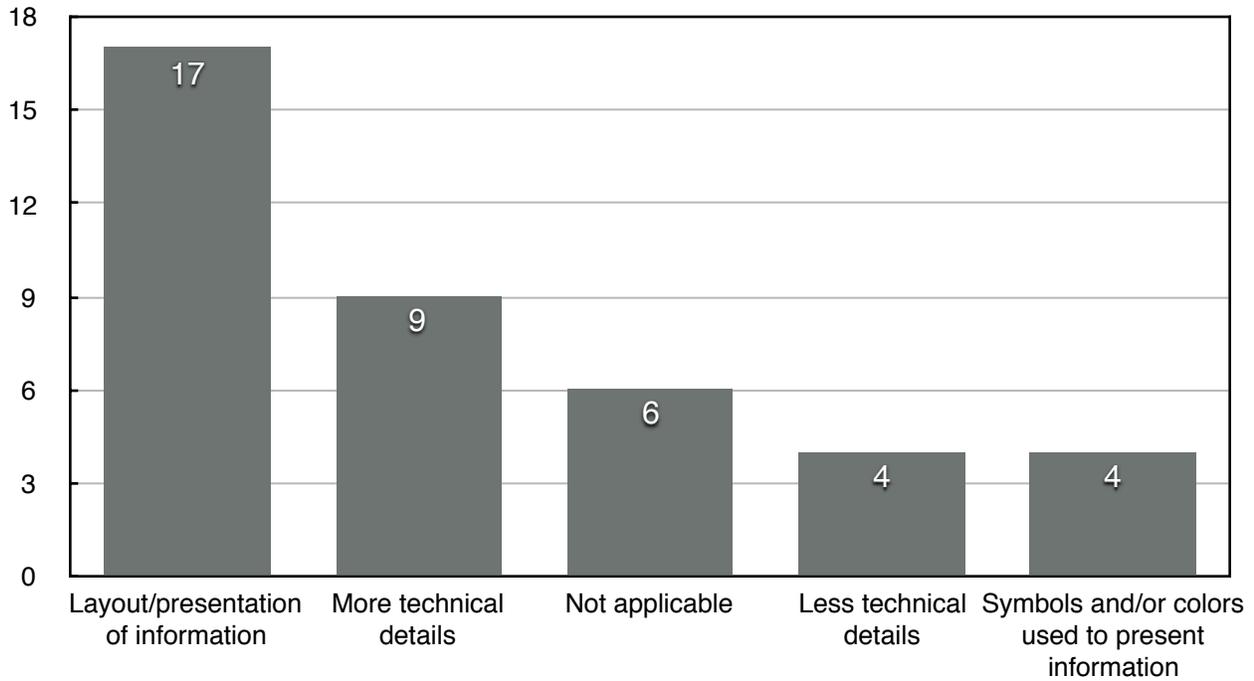

Which feature(s) of one vendor report makes it harder to interpret data compared to another vendor report? (n = 28, multiple answers allowed)

**Figure S3.8.**

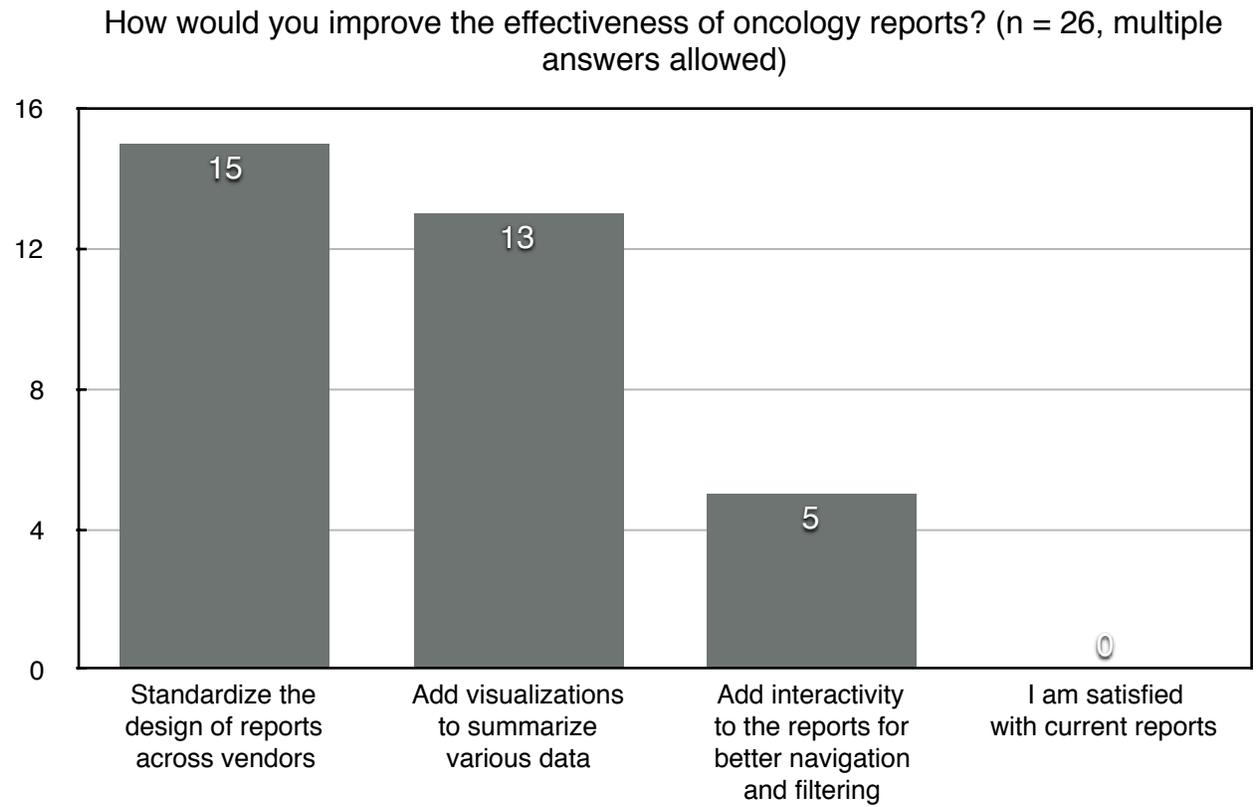

How would you improve the effectiveness of oncology reports? (n = 26, multiple answers allowed)

**Figure S3.9.**

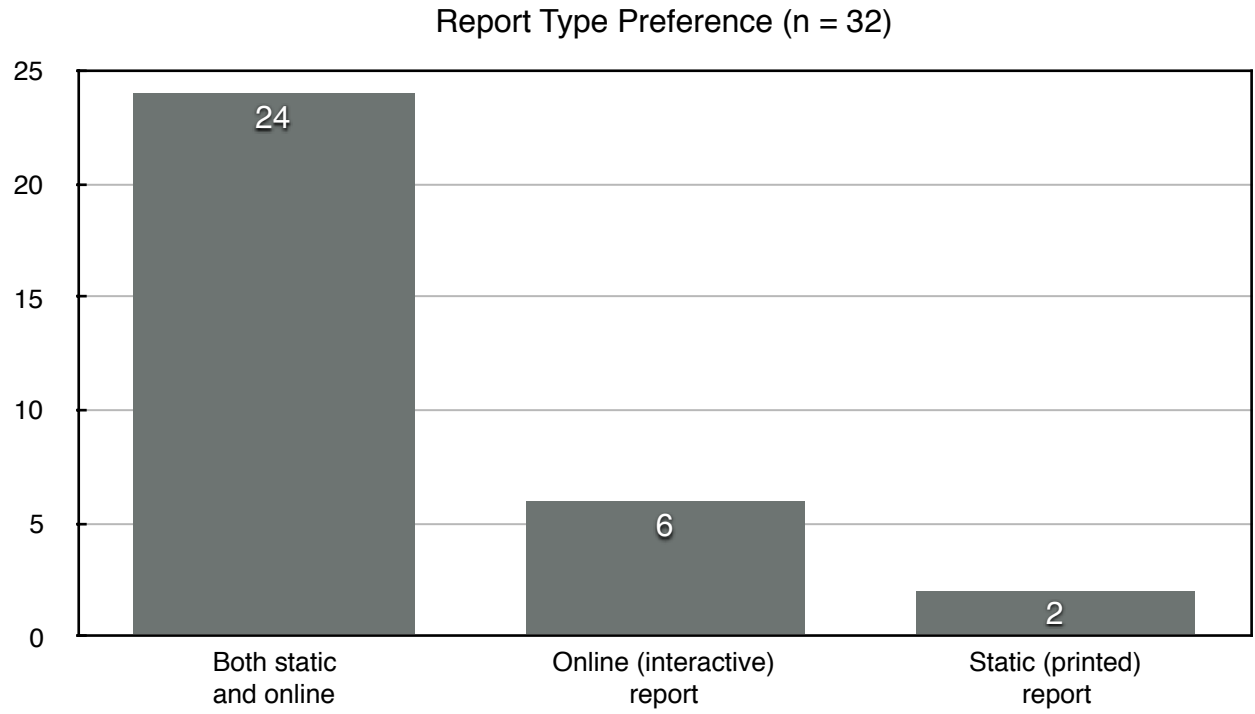

# SUPPLEMENTARY TABLE 1

# Precision Oncology Reports Survey

**Question 1-a:** *In the past 12 months, for what percentage of your patients did you order tumor genomic testing? Please include tests performed as part of a clinical trial. Your best estimate is fine.*

________ %

**Question 1-b:** *In the past 12 months, have you reviewed the following genomics test reports for your patients? Please check all that apply.*

- ☐ MSSM NGS Hotspot Cancer Panel
- ☐ BRAF
- ☐ EGFR
- ☐ KRAS
- ☐ MGMT
- ☐ MSI
- ☐ EML4-ALK
- ☐ C-KIT
- ☐ BCR-ABL

- ☐ JAK2
- ☐ PML-RARA
- ☐ MPL
- ☐ FLT3
- ☐ BCL2-IGH
- ☐ IGH rearrangement
- ☐ TRG rearrangement
- ☐ Other (please specify) ____________________

**Question 2:** *Which of the following vendor genomic test reports have you used? Please check all that apply.*

- ☐ FoundationOne
- ☐ Guardant360
- ☐ OncoVantage
- ☐ Caris Molecular Intelligence
- ☐ Pathway Genomics
- ☐ None
- ☐ Other (please specify) ________________

**Question 3:** *What is the most important information you expect to find in a tumor profiling report? Please check all that apply.*

- ☐   Genetic alterations/biomarkers
- ☐   Therapies
- ☐   Clinical trials
- ☐   Toxicity
- ☐   Other (please specify) __________________

**Question 4:** *What is the most time-consuming component of interpreting a tumor profiling report in your experience? Please check all that apply.*

- ☐   Actionable genetic alterations/biomarkers
- ☐   Therapies
- ☐   Clinical trials
- ☐   Toxicity
- ☐   Other (please specify) __________________

**Question 5:** *What parts are most troublesome reading in these reports? Please check all that apply.*

- ☐   Ease of locating the actionable summary
- ☐   Details of the actionable summary
- ☐   Verbosity of the technical information
- ☐   Ease of locating critical information
- ☐   Navigating cross-references in different sections of the report
- ☐   Other (please specify) __________________

**Question 6:** *Which feature(s) of one vendor report makes it harder to interpret data compared to another vendor report? Please check all that apply.*

- ☐ Less technical details
- ☐ More technical details
- ☐ Layout/presentation of information
- ☐ Symbols and/or colors used to present information
- ☐ Not applicable
- ☐ Other (please specify) _______________________________

**Question 7:** *How would you improve the effectiveness of oncology reports? Please check all that apply.*

- ☐ Add visualizations to summarize various data
- ☐ Add interactivity to the reports for better navigation and filtering
- ☐ Standardize the design of reports across vendors
- ☐ I am satisfied with current reports, no changes are necessary
- ☐ What do you suggest? _______________________________

**Question 8:** *Which report type would you prefer?*

- ☐ Static (printed) report
- ☐ Online (interactive) report
- ☐ Both static and online

**Question 9:** *How many years has it been since you graduated from medical school?*

- ☐ 0 - 5
- ☐ 6 - 10
- ☐ 11 - 15
- ☐ 16 - 20
- ☐ 21 - 25

- ☐ 26 - 30
- ☐ 31 - 35
- ☐ 36 - 40
- ☐ over 40 years

**Question 10:** *What is your gender?*

- ☐ Male
- ☐ Female

**Question 11:** *Are you a …*

- ☐ Surgical oncologist
- ☐ Medical oncologist
- ☐ Radiation oncologist
- ☐ Other (please specify) _______________________

**Question 12: *What is your clinical focus? Please check all that apply.***

- ☐ Blood cancer
- ☐ Breast cancer
- ☐ Cancer genetics and prevention
- ☐ Cutaneous cancer
- ☐ Gastrointestinal cancer
- ☐ Genitourinary cancer
- ☐ Gynecological cancer
- ☐ Head and neck cancer
- ☐ Hematology
- ☐ Melanoma
- ☐ Neuro-Oncology
- ☐ Sarcoma
- ☐ Thoracic cancer
- ☐ Other (please specify) ________________________

## COMMENTS

_______________________________________________________________

_______________________________________________________________

_______________________________________________________________

_______________________________________________________________

_______________________________________________________________

_______________________________________________________________